\documentclass[aps,prc,twocolumn,superscriptaddress]{revtex4-2}
\usepackage{CJKutf8}
\usepackage{amssymb}
\usepackage{amsmath}
\usepackage{diagbox}
\usepackage{makecell}
\usepackage{tikz}
\usepackage{color}
\usepackage{graphicx} 
\usepackage{multirow} 
\usepackage{makecell}
\usepackage{ulem}
\usepackage[hidelinks]{hyperref}

\begin{document}

\title{Octupole correlations in $^{220,222,224,226}$Rn}

\author{Yi-Ming Jiang}  
\affiliation{Institute of Theoretical Physics, Chinese Academy of Sciences, Beijing 100190, China}
\affiliation{School of Physical Sciences, University of Chinese Academy of Sciences, Beijing 100049, China} 
\author{Xiao Lu}  
\affiliation{Institute of Theoretical Physics, Chinese Academy of Sciences, Beijing 100190, China}
\author{Shan-Gui Zhou} 
\email[]{sgzhou@itp.ac.cn}
%\thanks{} 
\affiliation{Institute of Theoretical Physics, Chinese Academy of Sciences, Beijing 100190, China} 
\affiliation{School of Physical Sciences, University of Chinese Academy of Sciences, Beijing 100049, China} 
\affiliation{School of Nuclear Science and Technology, University of Chinese Academy of Sciences, Beijing 100049, China}  

\date{\today}

\begin{abstract}

\noindent \textbf{Background}: Radon isotopes are important candidates of observing octupole deformation. Octupole vibrational bands have been observed in $^{220,224,226}$Rn but no octupole deformation has been confirmed experimentally. %Morever, there are controversies in the predictions given by different models regarding whether radon isotopes exhibit octupole deformation.\\
\\
\textbf{Purpose}: To investigate whether octupole deformation exists in $^{220,222,224,226}$Rn and explore the possible octupole minima in the potential energy surfaces (PES). \\
\textbf{Methods}: The PES and ground-state properties are calculated by using the muti-dimensionally constrained relativistic mean-field (MDC-RMF) model with different covariant energy density functionals PC-PK1, DD-PC1, and DD-ME2. Pairing correlations are treated in the BCS approach with the separable pairing force of finite-range. \\
\textbf{Results}: The calculations show that the PES of all these isotopes exhibit softness with respect to octupole distortion and octupole minima appear in the PES of $^{222,224}$Rn with the above-mentioned functionals and several sets of pairing parameters used. The energy gain due to octupole deformation is found to be suppressed as the pairing strength increasing. As the octupole deformation increases, the neutron pairing energy decreases and the proton pairing energy increases while there is no obvious change in the total pairing energy. Analysis of the single-particle levels indicates that the observed octupole minima in $^{222,224}$Rn primarily originate from the coupling between neutron orbitals $\nu[642]3/2$ (from $\nu g_{9/2}$) and $\nu[761]3/2$ (from $\nu j_{15/2}$) near the Fermi surface. \\
\textbf{Conclusion}: The ground state properties and PES of $^{220,222,224,226}$Rn are studied and predictions are made that octupole minima exist in $^{222,224}$Rn. The energy gain due to octupole deformation is ranging from 0.08 to 0.79 MeV. Whether there is octupole deformation is still awaiting experimental confirmation, e.g., through measurement of the $E3$ transition strength from the ground
state to the first octupole $3^{-}$ state and observation of parity doublet bands with enhanced $E1$ transitions between them.
%The octupole correlations in $^{220,222,224,226}$Rn are investigated by using multi-dimensionally constrained covariant density functional theory. The ground-state properties and potential energy surfaces are analyzed, revealing that octupole deformation appears in $^{222,224}$Rn, but not in $^{220,226}$Rn. The relationship between pairing correlations and octupole deformation is examined, showing that the neutron pairing energy decreases as octupole deformation develops, whereas the proton pairing energy shows the opposite behavior. The microscopic origin of octupole correlations in these radon isotopes are explored based on an examination of the single-particle levels near the Fermi surface and a schematic two-level model. Experiments have indicated that these isotopes undergo octupole vibrations and the present prediction of octupole deformation in $^{222,224}$Rn awaits further confirmation.

\end{abstract}

%\keywords{}

\maketitle
\section{Introduction}

As a many-body quantum system, the atomic nucleus exhibits various  intrinsic shapes which are determined by the number of protons and neutrons and interactions among them. The axially deformed quadrupole deformation, assuming a reflection-symmetric nuclear shape, has been mostly studied since the 1950s. With certain numbers of protons and neutrons, some nuclei are expected to show reflection-asymmetric shapes, among which the octupole deformation was predicted many years ago \cite{Bohr1969_Nucl_Structure,LANE196039} and confirmed a decade ago \cite{Nature.497.199}. The octupole deformation plays important roles  not only in the ground-state properties and low-lying spectra \cite{RevModPhys.68.349,Butler_2016}, but also in the fission barriers of heavy and super-heavy nuclei \cite{PhysRevC.79.064304,Zhou_2016,PhysRevC.95.014303}. It is also crucial for the study of the intrinsic electric dipole moment \cite{PhysRevLett.114.233002,PhysRevLett.76.4316,PhysRevC.56.1357}.%, the EDM measurements of pear-shaped nuclei serve as a more sensitive probe for exploring the degree of charge parity breaking compared to the EDM of reflection symmetric nuclei \cite{PhysRevLett.114.233002}.  

When there are pairs of single-particle levels around the Fermi surface with opposite parity and $\Delta l=\Delta j=3\hbar$ ($l$ represents orbital angular momentum and $j$ total angular momentum), octupole correlations arise in the nucleus and these correlations, if strong enough, may induce a pear-shaped nucleus \cite{RevModPhys.68.349}. This condition is met when the proton number is around 34, 56, and 88, and the neutron number is around 34, 56, 88, and 134 \cite{NAZAREWICZ1984269,PhysRevC.41.R1351,PhysRevC.42.1264}.
%Besides, The existence of octupole deformation has been experimentally confirmed \cite{PhysRevLett.118.152504,PhysRevLett.116.112503,Nature.497.199,Nature.497.199,Nature.497.199}. 
%The region of $A \approx220$ in which the proton number is about 88 and neutron number is about 134 is considered to be a island of octupole correlation and octupole deformation. %Experimentally, static octupole deformation have been discover   ed in this region,  such as in $^{222}$Ra \cite{Nature.497.199}, $^{224}$Ra \cite{PhysRevLett.124.042503} and $^{228}$Th \cite{Nature.497.199}. Another candidate in this island is radon with proton number 86 and several experimental explorations have been performed. Currently, although static octupole deformation hasn't been seen in radon isotopes, ${}^{220}$Rn, ${}^{224}$Rn and ${}^{226}$Rn undergo octupole vibrational bands \cite{Nature.497.199,Nat.Commun.10.2403}. However, whether a stable octupole deformation exists throughout the Rn isotope chain remains unknown.
Octupole deformation or octupole correlations have been experimentally observed in the mass regions $A \sim$ 80, 150, and 220 \cite{PhysRevLett.116.112501,doi:10.1142/S0218301323400037,PhysRevLett.116.112503,Nature.497.199,PhysRevLett.124.042503,Nat.Phys.16.853,Nat.Commun.10.2403}. In particular, in the $A\sim$ 220 ($Z\sim$ 88, $N\sim$ 134) mass region, such as $^{222}$Ra \cite{Nature.497.199}, $^{224}$Ra \cite{PhysRevLett.124.042503}, and $^{228}$Th \cite{Nat.Phys.16.853} are recognized as exhibiting typical octupole deformation, and this region is referred to as one of the islands of octupole deformation. Within this island, radon isotopes are also considered as candidates for octupole deformation, with octupole vibrational bands observed in $^{220}$Rn \cite{Nature.497.199} and $^{224,226}$Rn \cite{Nat.Commun.10.2403}. %\sout{However, whether octupole deformation exists in the radon isotope chain remains unclear.}
%\red{Nevertheless, the same study on $^{224,226}$Rn \cite{Nat.Commun.10.2403} concluded that these isotopes ‘do not possess static pear-shapes’. This is in contrast to theoretical predictions from macro-microscopic models \cite{NAZAREWICZ1984269,MOLLER20161}, which suggest the presence of octupole deformation in $^{224,226}$Rn. Such conflicting conclusions, both experimental and theoretical, underscore the necessity for a systematic and theoretical study to clarify the situation in radon isotopes.} \blue{(XLU: precise不建议用)}

Theoretically, many approaches have been developed to describe octupole correlations in nuclei, such as the macro-microscopic model \cite{NAZAREWICZ1984269,PhysRevC.95.034329},  Hartree-Fock-Bogoliubov  (HFB)  theories with Gogny interaction \cite{PhysRevC.84.054302,Robledo_2012,Robledo_2015} and Skyrme interactions \cite{PhysRevC.85.025802,Ebata_2017,PhysRevC.102.024311,PhysRevC.108.024303}, relativistic mean field (RMF) model \cite{Zhou_2016,PhysRevC.102.024311,LI2013866,RONG2023137896,XU2024138893,LU2025139620}, cranked shell model \cite{NAZAREWICZ1985420,PhysRevC.102.064328}, projected shell model \cite{PhysRevC.63.014314,PhysRevC.91.014317}, cluster model \cite{PhysRevC.67.014313,Buck2008NegativePB,PhysRevC.92.034302}, particle rotor model \cite{LEANDER1984375,WANG2019454}, and interacting boson model \cite{ENGEL198761,PhysRevC.58.1500,PhysRevC.67.014305}. However, there are controversies in the predictions given by different models regarding whether radon isotopes exhibit octupole deformation. Prediction using macro-microscopic model with a Woods-Saxon potential has been made that there is octupole deformation in $^{222}$Rn and $^{224}$Rn \cite{NAZAREWICZ1984269}; calculations using the finite-range liquid-drop model have shown that $^{220,222,224,226}$Rn all exhibit octupole deformation \cite{MOLLER20161}; HFB calculations with the Gogny interaction have predicted octupole deformation in $^{220}$Rn at the mean field level \cite{PhysRevC.86.054306,PhysRevC.88.051302}. With some other models, no octupole deformation has been revealed in radon isotopes \cite{NAZAREWICZ1984269,PhysRevC.93.044304}.

The multi-dimensionally constrained covariant density functional theory (MDC-CDFT) \cite{PhysRevC.89.014323,Zhou_2016,PhysRevC.85.011301,PhysRevC.95.014320} have already been developed to study various shapes and their manifestations in nuclei. In MDC-CDFT, all deformations  governed by the $V_{4}$ symmetry, including $\beta_{\lambda\mu}$ with even $\mu$ are self-consistently considered. With different methods to deal with pairing correlations, two variants have been developed: the MDC mean-field model (MDC-RMF) with the BCS (Bardeen-Cooper-Schrieffer) approach \cite{PhysRevC.85.011301,PhysRevC.89.014323}, and the MDC relativistic Hartree–Bogoliubov model (MDC-RHB) with the Bogoliubov transformation \cite{PhysRevC.95.014320}. The MDC-CDFT has been successfully used in investigating fission barriers of actinide nuclei \cite{PhysRevC.89.014323,PhysRevC.85.011301}, the third minima in potential energy surfaces (PESs) of light actinides \cite{PhysRevC.91.014321}, shapes and PESs of superheavy nuclei \cite{Meng:2019mff,Wang_2022}, nonaxial octupole $Y_{32}$ correlations \cite{PhysRevC.95.014320,PhysRevC.86.057304}, the structure of hypernuclei \cite{PhysRevC.84.014328,PhysRevC.89.044307,RONG2020135533,Chen:2021kde,Sun_2022}, and the low-lying excited states associated with exotic nuclear shapes \cite{Wang_2022_c,RONG2023137896,LU2025139620}. In this work, we use MDC-RMF to investigate radon isotopes. Calculations are performed to study octupole correlations and other ground-state properties in $^{220,222,224,226}$Rn and to explain the microscopic origin of octupole deformation. 

The paper is organized as follows. Section \ref{TF} presents a review of the theoretical framework. The results and discussions are presented in Sec. \ref{RD}. A brief summary and perspectives are given in Sec. \ref{SP}.  

\section{Theoretical framework}\label{TF}

The covariant density functional theory (CDFT), self-consistently including the relativistic effects and giving an effective description of nuclear many-body systems, has received wide attention due to its successful description of nuclear phenomena during the past years \cite{Zhou_2016,serot1986,Reinhard_1989,RING1996193,VRETENAR2005101,MENG2006470,Paar_2007,NIKSIC2011519}. In CDFTs, either meson exchange (ME) or point coupling (PC) functionals can be used. Furthermore, the couplings can be nonlinear (NL) or density-dependent (DD).  In this section, we only briefly review the RMF model with nonlinear point-coupling (NL-PC). %Detailed descriptions can be found in Refs.  \cite{BOGUTA1977413,PhysRevLett.68.3408,SUGAHARA1994557}. 

The Lagrangian density of the RMF model with NL-PC \cite{PhysRevLett.68.3408,Sun_2022,PhysRevC.46.1757,PhysRevC.65.044308} can be written as
\begin{equation}
\mathcal{L}=\bar{\psi}(i\gamma_\mu\partial^\mu-M)\psi-\mathcal{L}_{\rm{4f}}-\mathcal{L}_{\rm{nl}}-\mathcal{L}_{\rm{der}}-\mathcal{L}_{\rm{em}},
\end{equation} 
where
\begin{equation}
	\begin{aligned}
		\mathcal{L}_{\rm{4f}}=\ &\frac{1}{2}\alpha_S\rho^2_S+\frac{1}{2}\alpha_V\rho^2_V+\frac{1}{2}{\alpha}_{TS}\vec{\rho}^2_{TS}+\frac{1}{2}{\alpha}_{TV}\vec{\rho}^2_{TV},\\
		\mathcal{L}_{\rm{nl}}=\ &\frac{1}{3}\beta_S\rho^3_S+\frac{1}{4}\gamma_S\rho^4_S+\frac{1}{4}\gamma_V[\rho^2_V]^2,\\
		\mathcal{L}_{\rm{der}}=\ &\frac{1}{2}\delta_S[\partial_\nu\rho_S]^2+\frac{1}{2}\delta_V[\partial_\nu\rho_V]^2+\frac{1}{2}\delta_{TS}[\partial_\nu\vec{\rho}_{TS}]^2\\
		&+\frac{1}{2}\delta_{TV}[\partial_\nu\vec{\rho}_{TV}]^2,\\
		\mathcal{L}_{\rm{em}}=\ &\frac{1}{4}F^{\mu\nu}F_{\mu\nu}+e\frac{1-\tau_3}{2}A_0\rho_V\\
	\end{aligned}
\end{equation}
are linear coupling, nonlinear coupling, derivative coupling, and electromagnetic interaction part, respectively. $M$ represents nucleon mass and $\rho_S, \rho_V, \vec{\rho}_{TS}$, and $\vec{\rho}_{TV}$ are isoscalar density, isoscalar current, isovector density, and isovector current, respectively. $\alpha_S$, $\alpha_V$, $\alpha_{TS}$, $\alpha_{TV}$, $\beta_{S}$, $\gamma_{S}$, $\gamma_V$, $\delta_S$, $\delta_V$, $\delta_{TS}$, and $\delta_{TV}$ are coupling constants of different channels and $e$ represents the electric charge. In a system with time-reversal symmetry, only time-like components contribute, so these densities and currents can be written as
\begin{equation}
    \label{3}
	\begin{aligned}
		&\rho_S=\bar{\psi}\psi, \\
       & \rho_V=\bar{\psi}\gamma_0\psi,\\
		&\vec{\rho}_{TS}=\bar{\psi}\vec{\tau}\psi, \\
&\vec{\rho}_{TV}=\bar{\psi}\gamma_0\vec{\tau}\psi.\\
	\end{aligned}
\end{equation}
From this Lagrangian, with Hartree approximation and no sea approximation, one can get the equation of motion for a nucleon
\begin{equation}
	\hat{h}\psi_k(\vec{\mathbf{r}})=\varepsilon_k\psi_k(\vec{\mathbf{r}}),
\end{equation}
where the single-particle Hamiltonian $\hat{h}$ can be written as
\begin{equation}
	\hat{h}=\mathbf{\alpha\cdot p}+\beta(M+S(\vec{\mathbf{r}}))+V(\vec{\mathbf{r}}),
\end{equation}
where the scalar potential $S(\mathbf{r})$ and vector potential $V(\mathbf{r})$ read
\begin{equation}
\label{6}
     \begin{aligned}
     	S(\vec{\mathbf{r}})=\ &\alpha_S\rho_S+\alpha_{TS}\vec{\rho_{TS}}\cdot\vec{\tau}+\beta_S\rho^2_S+\gamma_S\rho^3_S\\
     	&+\delta_S\Delta\rho_S+\delta_{TS}\Delta\vec{\rho}_{TS}\cdot\vec{\tau},\\
     	V(\vec{\mathbf{r}})=\ &\alpha_V\rho_S+\alpha_{TV}\vec{\rho_{TV}}\cdot\vec{\tau}+\gamma_V\rho^3_V\\
     	&+\delta_V\Delta\rho_V+\delta_{TV}\Delta\vec{\rho}_{TV}\cdot\vec{\tau}.\\
     \end{aligned}
\end{equation}

In MDC-RMF, the BCS method is used to deal with pairing correlations in finite nuclei. In the present work, we use a separable pairing force of finite-range \cite{TianYuan_2006,TIAN200944,PhysRevC.79.064301,PhysRevC.80.024313}
\begin{equation}
	V=-G\delta(\mathbf{R-R^\prime})P(r)P(r^\prime)\frac{1-P^\sigma}{2},
\end{equation}
where $\mathbf{R}=(\mathbf{r_1+r_2})/2$ and $\mathbf{r=r_1-r_2}$ are center-of-mass and relative coordinates, respectively. The factor $P(r)$ is a Gaussian function
\begin{equation}
	P(r)=\frac{1}{(4\pi a^2)^{3/2}}\exp{\left(-\frac{r^2}{4a^2}\right)}.
\end{equation}
The values of pairing strength $G$ and effective range of the pairing force $a$ are fitted to the ${}^{1}\rm{S}_0$ channel pairing gap of nuclear matter with the Gogny force. The separable pairing force of finite-range can be easily applied in realistic applications of modern relativistic and non-relativistic density functional theory \cite{PhysRevC.80.024313,TIAN200944}. Two sets of parameters are used in this work: $G=G_0=738$ MeV fm${}^{3}$ and $a=0.636$ fm for Gogny force D1, and $G=G_0=728$ MeV fm${}^{3}$ and $a=0.644$ fm for Gogny force D1S. In the following, we denote these two sets of pairing parameters with ``S-D1'' and ``S-D1S''.
\begin{figure*}[tbp] % [t]表示优先放在页面顶部
    \centering
    \includegraphics[width=\textwidth]{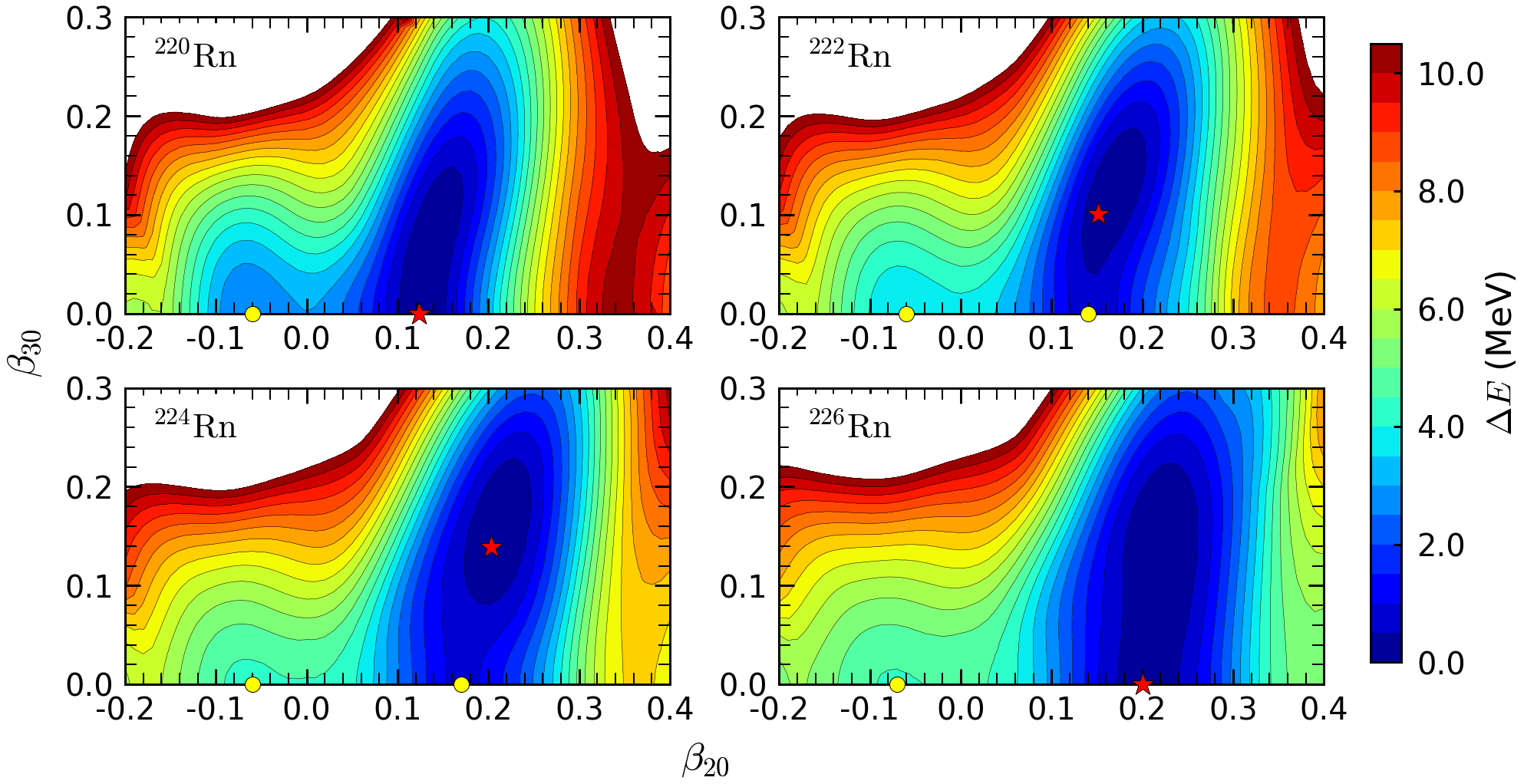} % 使用\textwidth自动适应通栏宽度
    \caption{The potential energy surfaces of $^{220,222,224,226}$Rn in the $\beta_{20}$--$\beta_{30}$ plane from MDC-RMF calculations by using parameter set DD-ME2$+$S-D1. The energy is normalized with respect to the binding energy of the ground state. The counter interval is 0.5 MeV. The ground state and other local minima are marked by red stars and yellow dots, respectively.}
    \label{fig:pes}
\end{figure*}

In our study, the well-used and successful functionals PC-PK1 \cite{PhysRevC.82.054319,PhysRevC.86.064324}, DD-PC1 \cite{PhysRevC.78.034318}, and DD-ME2 \cite{PhysRevC.71.024312} are chosen to ensure reliable conclusions and check the functional-dependence of our results. The equations of motion are solved in the axially-deformed harmonic oscillator (ADHO) basis \cite{GAMBHIR1990132,RING199777}. For constrained calculations, a modified linear constrain of multipole moment are used
\begin{equation}
    E^\prime = E_{\rm{RMF}} + \sum\limits_{\lambda\mu} {\frac{1}{2}C_{\lambda\mu}\beta_{\lambda\mu}},
\end{equation}
where the variable $C_{\lambda\mu}$ varies during the iteration. The deformation parameters $\beta_{\lambda\mu}$ are defined as
\begin{equation}
	\beta_{\lambda\mu}=\frac{4\pi}{3AR^\lambda}\langle \hat{Q}_{\lambda\mu} \rangle,
\end{equation}
where $R=1.2A^{1/3}$ fm is the nuclear radius, and $A$ represents the mass number of the nucleus. The multipole moment operator is
\begin{equation}
	\hat{Q}_{\lambda\mu}=r^\lambda Y_{\lambda\mu}(\theta,\phi),
\end{equation}
$Y_{\lambda\mu}$ is spherical harmonic function. This method achieves multidimensional constraints in deformation space and is referred to as MDC-CDFT, allowing us to calculate PESs and investigate nuclear ground state and fission properties.

\section{Results and discussions}\label{RD}

\subsection{Numerical Details}

To investigate the PESs and ground-state properties, the functionals PC-PK1, DD-PC1, and DD-ME2, and pairing force parameters ``S-D1'' and ``S-D1S', are chosen to perform MDC-RMF calculations.  For the following study, we assume these nuclei have axially symmetric shapes. 

The truncation of the basis is performed that all states belonging to the major shells up to $N_{\rm{F}}=20$ fermionic shells for the Dirac spinors and up to $N_{\rm{B}}=26$ bosonic shells for the meson fields are taken into account. This truncation scheme has been numerically verified for light actinides   \cite{PhysRevC.85.011301,PhysRevC.89.014323}. The basis deformation $\beta_{ \text{basis}}$  is chosen in the following way: $\beta_{\text{basis}} = \beta_{20}$ for $\beta_{20}<0.3$ and $\beta_{\text{basis}} = \beta_{20}/2$ for $0.3<\beta_{20}<0.4$. 

The various densities (\ref{3}) and the scalar and vector potentials (\ref{6}) are calculated on a spatial lattice, where the number of mesh points in the $\rho$ and $z$ directions is set to 36 and 18, respectively, ensuring the convergence of the calculation results. The PESs are obtained through constrained calculations in the $\beta_{20}$--$\beta_{30}$ plane, with the $\beta_{20}$ values ranging from $-0.2$--0.4 and the $\beta_{30}$ values ranging from 0.0--0.3 with a deformation step of 0.01 in each direction.

\subsection{PESs and Ground-State Properties }

 Figure \ref{fig:pes} displays the PESs for ${}^{220,222,224,226}$Rn obtained by using the parameter set DD-ME2+S-D1. The PES of ${}^{220}$Rn indicates that its ground state exhibits quadrupole deformation but no octupole deformation. With the neutron number increasing, the non-octupole deformed minimum becomes a local minimum, and the global minima (ground states) of ${}^{222,224}$Rn shift towards the octupole direction, leading to octupole-deformed ground states. For ${}^{226}$Rn, this octupole-deformed minimum disappears, and the quadrupole-deformed minimum reappears as the ground state. Moreover, a local minimum of oblate deformation consistently exists across all isotopes. However, its energy remains significantly higher than other minima, precluding it from becoming the ground state. Furthermore, the PESs show that these nuclei are soft with respect to the octupole distortion, likely linked to the experimentally observed vibrational bands. Calculations with other parameter sets give similar results. 
\begin{table}[htbp]  
 \centering
 \caption{Ground-state properties of ${}^{220,222,224,226}$Rn, including binding energy $E_{\rm{bind}}$, quadrupole deformation $\beta_{20}$, octupole deformation $\beta_{30}$, and the energy gain of octupole deformation $\Delta E_{\rm{oct}}$ [defined in Eq. (\ref{eq})]. The experimental data (Exp.) for binding energy are taken from Refs. \cite{Huang_2021,Wang_2021}. $E_{\rm{bind}}$, $\Delta E_{\rm{oct}}$, and experimental data are all in MeV.}\label{tab1}
 \begin{tabular}{p{1.1cm} p{2.5cm} p{1.3cm} p{1.0cm} p{1.0cm} p{1.0cm}}
 \hline
 \hline
 Nucleus & Parameter set & $E_{\rm{bind}}$ & $\beta_{20}$ & $\beta_{30}$ & $\Delta E_{\rm{oct}}$\\
 \hline
 ${}^{220}$Rn & DD-PC1+S-D1S & 1697.96 & 0.122 & 0.000 & 0.00 \\
              & DD-PC1+S-D1  & 1698.04 & 0.119 & 0.000 & 0.00 \\
              & PC-PK1+S-D1S & 1696.52 & 0.127 & 0.000 & 0.00 \\
              & PC-PK1+S-D1  & 1696.71 & 0.125 & 0.000 & 0.00 \\
              & DD-ME2+S-D1S & 1697.22 & 0.124 & 0.000 & 0.00 \\
              & DD-ME2+S-D1  & 1697.29 & 0.124 & 0.000 & 0.00 \\
              & Exp.         & 1697.80 &       &       &      \\
 ${}^{222}$Rn & DD-PC1+S-D1S & 1707.96 & 0.154 & 0.099 & 0.29 \\
              & DD-PC1+S-D1  & 1708.06 & 0.156 & 0.101 & 0.23 \\
              & PC-PK1+S-D1S & 1706.81 & 0.162 & 0.102 & 0.12 \\
              & PC-PK1+S-D1  & 1707.01 & 0.160 & 0.097 & 0.08 \\
              & DD-ME2+S-D1S & 1707.12 & 0.154 & 0.095 & 0.53 \\
              & DD-ME2+S-D1  & 1706.25 & 0.157 & 0.101 & 0.78 \\
              & Exp.         & 1708.17 &       &       &      \\
 ${}^{224}$Rn & DD-PC1+S-D1S & 1717.33 & 0.183 & 0.105 & 0.19 \\
              & DD-PC1+S-D1  & 1717.52 & 0.183 & 0.104 & 0.16 \\
              & PC-PK1+S-D1S & 1716.45 & 0.198 & 0.136 & 0.13 \\
              & PC-PK1+S-D1  & 1716.61 & 0.183 & 0.127 & 0.12 \\
              & DD-ME2+S-D1S & 1716.64 & 0.203 & 0.139 & 0.52 \\
              & DD-ME2+S-D1  & 1716.79 & 0.204 & 0.140 & 0.79 \\
              & Exp.         & 1718.16 &       &       &      \\
 ${}^{226}$Rn & DD-PC1+S-D1S & 1726.85 & 0.194 & 0.000 & 0.00 \\
              & DD-PC1+S-D1  & 1727.05 & 0.195 & 0.000 & 0.00 \\
              & PC-PK1+S-D1S & 1726.06 & 0.197 & 0.000 & 0.00 \\
              & PC-PK1+S-D1  & 1726.26 & 0.197 & 0.000 & 0.00 \\
              & DD-ME2+S-D1S & 1726.09 & 0.197 & 0.000 & 0.00 \\
              & DD-ME2+S-D1  & 1726.24 & 0.200 & 0.000 & 0.00 \\
              & Exp.         & 1728.09 &       &       &      \\
 \hline 
 \hline
 \end{tabular}
\end{table}
To obtain a more accurate description of the ground state, unconstrained self-consistent calculations are performed near the minima of the PESs. The calculated ground-state properties are listed in  Table \ref{tab1}. The systematically calculated binding energies by using various parameter sets are very close to the experimental data. The results with all these parameter sets predict no octupole deformation for $^{220,226}$Rn, while they consistently indicate octupole deformation in $^{222}$Rn with $\beta_{30}\sim0.1$ and $^{224}$Rn with $\beta_{30}\sim$ 0.1--0.14. 

To quantify the impact of octupole deformation on the ground-state binding energy, we introduce the energy gain of octupole deformation $\Delta E_{\rm{oct}}$, defined as follows,
\begin{equation}
      \Delta E_{\rm{oct}}=E_{\rm{oct}}-E_{\rm{quad}},
      \label{eq}
\end{equation}
where $E_{\rm{quad}}$ ($E_{\rm{oct}}$) is the binding energy obtained from MDC-RMF calculations with reflection symmetry imposed (released). As can be seen in  Table \ref{tab1}, the energy gains of octupole deformation in $^{222}$Rn and $^{224}$Rn are different across different parameter sets. The smallest energy gain of $\Delta E_{\rm{oct}}\sim$ $0.1$ MeV is obtained with PC-PK1, and slightly larger values are produced with DD-PC1. The calculations with DD-ME2 predict greater octupole stabilization, with $\Delta E_{\rm{oct}}\sim$ $0.5$ MeV with S-D1S and $\Delta E_{\rm{oct}}$ up to $\sim$ $0.8$ MeV with S-D1. %\red{The $\Delta E_{\rm{oct}}$ is ranging from 0.08 to 0.79 MeV and whether there is octupole deformation is still awaiting experimental confirmation.}

\subsection{The Pairing Effects}

It is known that pairing correlations play an important role in binding energies. Generally, pairing correlations counteract shell effects, having a tendency to make the system less deformed \cite{RevModPhys.68.349}. In this section, we vary the pairing strength parameter $G$ by 10\% to investigate energy gains of octupole deformation and examine the variation of pairing energy with octupole deformation.

\begin{figure}[htbp] % [t]表示优先放在页面顶部
    \centering
    \includegraphics[width=0.40\textwidth]{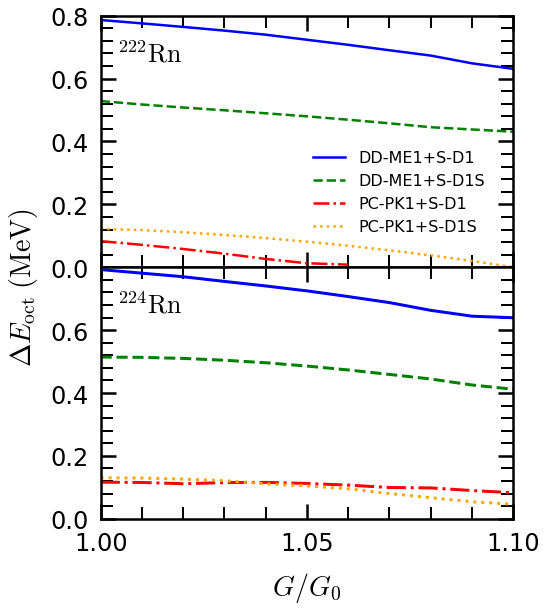} % 使用\textwidth自动适应通栏宽度
    \caption{The energy gains of octupole deformation $\Delta E_{\rm{oct}}$ [defined in Eq. (\ref{eq})]  of $^{222,224}$Rn as a function of pairing strength $G$ scaled by $G_0$. The calculated results with parameter sets DD-ME2+S-D1 (blue solid line), DD-ME2+S-D1S (green dash line), PC-PK1+S-D1 (red dash-dotted line), and PC-PK1+S-D1S (orange dotted line) are presented.}
    \label{fig:E}
\end{figure}
\begin{figure*}[htbp]
    \centering
    \includegraphics[width=1.0\linewidth]{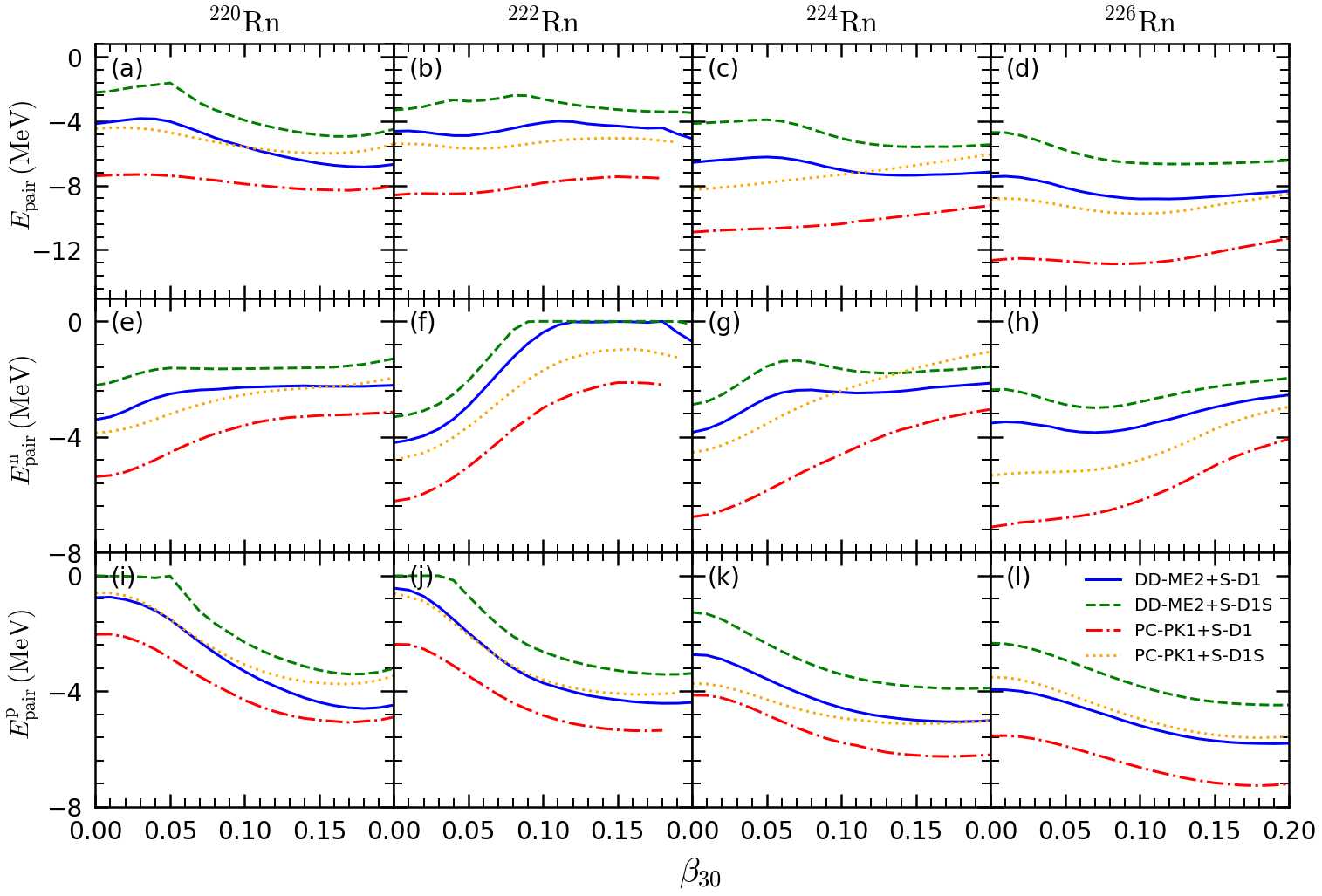} % 调整为单栏宽度 (0.8倍线宽)
    \caption[Pairing energy versus octupole deformation]{The total pairing energies $E_{\rm{pair}}$ (a)\color{black}--(d),  neutron pairing energies $E^{\rm{n}}_{\rm{pair}}$ (e)\color{black}--(h), and proton pairing energies $E^{\rm{p}}_{\rm{pair}}$ (i)\color{black}--(l) of $^{220,222,224,226}$Rn as a function of octupole deformation $\beta_{30}$. The calculated results with parameter sets DD-ME2+S-D1 (blue solid line), DD-ME2+S-D1S (green dash line), PC-PK1+S-D1 (red dash-dotted line), and PC-PK1+S-D1S (orange dotted line) are shown. The quadrupole deformations $\beta_{20}$ of these nuclei are fixed at the ground-state values as given in Table \ref{tab1}. 
    }
    \label{fig:pair}
\end{figure*}

Figure \ref{fig:E} displays the energy gains of octupole deformation $\Delta E_{\rm{oct}}$  as a function of pairing strength parameter $G$ scaled by $G_0$ using different parameter sets. These results reveal a universal trend: Octupole energy gain $\Delta E_{\rm{oct}}$ decreases monotonically as the pairing strength increases, demonstrating the suppression role of pairing on octupole deformation. Results obtained with PC-PK1 functional exhibits a particularly different behavior in $^{222}$Rn: when the pairing strength becomes larger, $\Delta E_{\rm{oct}}$ vanishes, signaling a transition from octupole-deformed to non-octupole-deformed ground state. %These observations align with the well-established understanding that pairing interactions tend to suppress the emergence of octupole deformation.

\begin{figure*}[htbp] % [htbp]表示优先放在页面顶部
    \centering
    \includegraphics[width=0.9\textwidth]{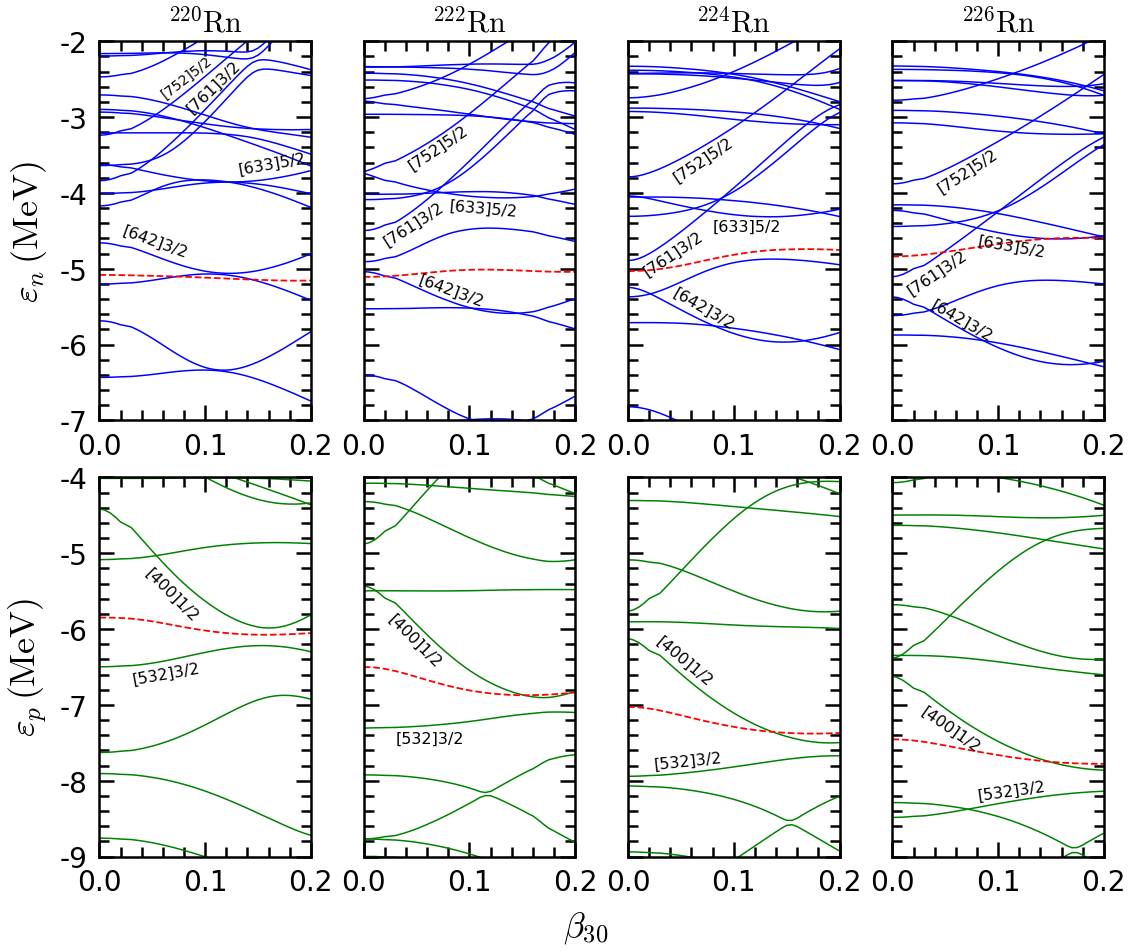} % 使用\textwidth自动适应通栏宽度
    \caption{Single-particle levels of $^{220,222,224,226}$Rn as a function of octupole deformation $\beta_{30}$ calculated with parameter set DD-ME2+S-D1. The quadrupole deformations $\beta_{20}$ of these nuclei are fixed at the ground-state values as given in Table \ref{tab1}. Blue solid lines, green solid lines, and red dash lines represent single-neutron levels, single-proton levels, and Fermi levels, respectively. For convenience, single-neutron orbitals strongly coupled by octupole interaction and single-proton orbitals strongly perturbed by octupole deformation around the Fermi surface are marked with Nilsson quantum numbers at $\beta_{30}=0$.}
    \label{fig:level}
\end{figure*}

Figure \ref{fig:pair} shows the total pairing energies $E_{\rm{pair}}$, neutron pairing energies $E^{\rm{n}}_{\rm{pair}}$, and proton pairing energies $E^{\rm{p}}_{\rm{pair}}$ of $^{220,222,224,226}$Rn as a function of octupole deformation $\beta_{30}$. It is clearly seen that the proton pairing energy $E^{\rm{pair}}_{\rm{p}}$ increases with octupole deformation $\beta_{30}$, while the neutron pairing energy $E^{\rm{n}}_{\rm{pair}}$ decreases with octupole deformation $\beta_{30}$. As a result, the total pairing energy $E_{\rm{pair}}$ stays nearly constant. 

The behavior of pairing energy with octupole deformation is related to the single-particle levels near the Fermi surface. Figure ~\ref{fig:level} shows single-particle levels of $^{220,222,224,226}$Rn as a function of octupole deformation $\beta_{30}$, with single-particle orbitals strongly affected by octupole interactions around the Fermi surface marked by their corresponding Nilsson quantum numbers at $\beta_{30}=0$. As shown in Fig.~\ref{fig:level}, the single-proton levels near the Fermi surface move closer as the octupole deformation $\beta_{30}$ increases, such as single-proton levels originating from Nilsson levels $\nu$[400]1/2 and $\nu$[532]3/2, leading to enhanced pairing effects. Conversely, single-neutron levels near the neutron Fermi surface move away from the Fermi surface with increasing octupole deformation, such as single-neutron levels originating from Nilsson levels $\nu$[642]1/2 and $\nu$[761]3/2, weakening pairing effects and reducing neutron pairing energy. As these orbitals are not so close to the Fermi level in $^{220}$Rn, the decreasing trend of neutron pairing energy is not significant. %In summary, one cannot simply assume that octupole deformation invariably reduces pairing energies, as single-particle  levels play a pivotal role in mediating this relationship. 

\subsection{The Microscopic Origin of Octupole Deformation}

To understand why octupole minima exist in $^{222,224}$Rn but are absent in $^{220,226}$Rn, we examine the single-particle structures of these nuclei. In nuclei with $A \approx 220$, the emergence of octupole deformation is primarily determined by the couplings between neutron orbitals $\nu g_{9/2}$$\leftrightarrow$$\nu j_{15/2}$ and between proton orbitals $\pi f_{7/2}$$\leftrightarrow$$\pi i_{13/2}$. When these orbitals locate near the Fermi surface, they may induce octupole-deformed shapes in nuclei. For the isotopic chain of radon investigated in this work, as evidenced in the lower pannels of Fig.~\ref{fig:level}, the single-proton levels in $^{220,222,224,226}$Rn show similar behaviors as $\beta_{30}$ develops. In addition, proton orbitals with strong octupole correlations are far away from the Fermi surface, e.g., orbitals originating from $\pi f_{7/2}$ are around $-$12 MeV and not shown in Fig.~\ref{fig:level}. Therefore, the octupole correlations of proton orbitals should not be the main origin of octupole deformation in $^{222,224}$Rn. To unravel the origin of octupole deformation in these two nuclei, next we make an analysis of single-neutron levels near the Fermi surface.      

\begin{figure}[htbp] % [htbp]表示优先放在页面顶部
    \centering 
    \includegraphics[width=0.44\textwidth]{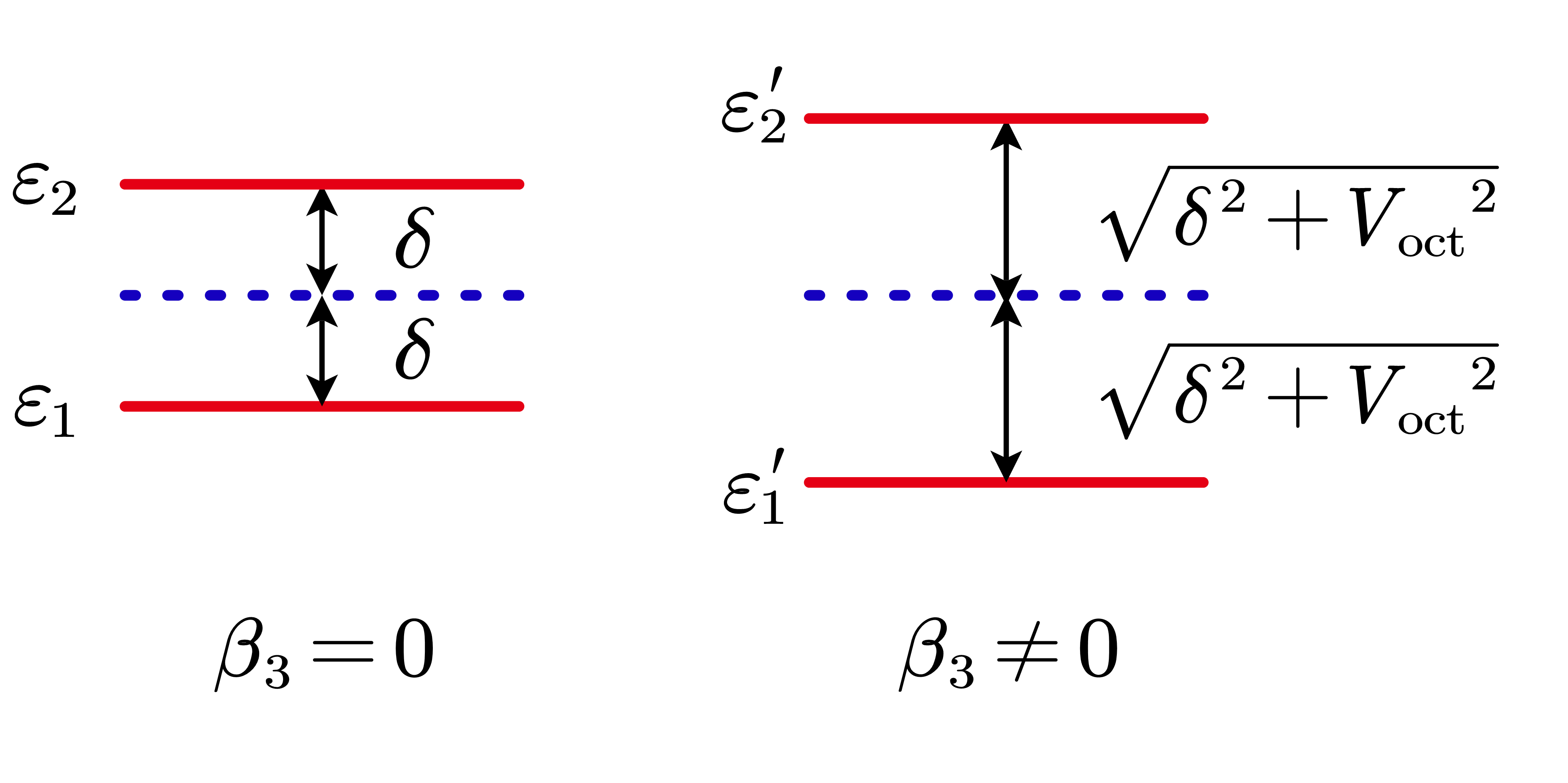} % 使用\textwid  th自动适应通栏宽度
    \caption{Evolution of a pair of single-particle orbitals perturbed by octupole interaction. The energies of two single-particle levels are marked as $\varepsilon_{1}$ and $\varepsilon_{2}$ (without octupole deformation) or $\varepsilon^\prime_{1}$ and $\varepsilon^\prime_{2}$ (with octupole deformation). $2\delta$ represents the energy gap at $\beta_{30}=0$. $V_{\rm{oct}}$ represents the off-diagonal element of the single-particle Hamiltonian with octupole interaction.}% Orbitals coupled by octupole interaction split in the reflection-asymmetric mean field, while the total energy of these two orbitals remains conserved.}
    \label{fig:oct}
\end{figure}
\begin{figure*}[htbp] % [htbp]表示优先放在页面顶部
    \centering
    \includegraphics[width= 0.7\textwidth]{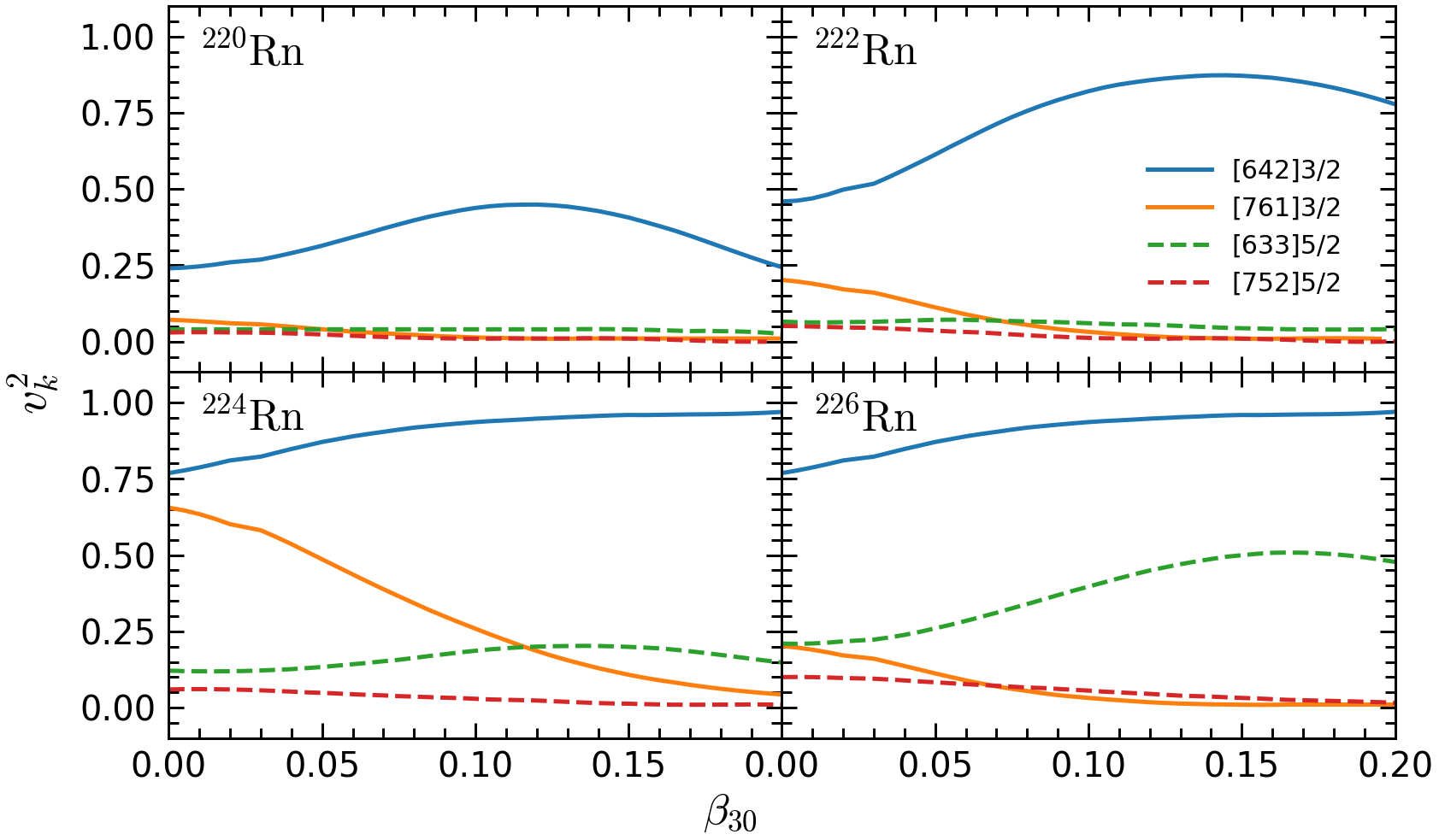} % 使用\textwidth自动适应通栏宽度
    \caption{Occupation probabilities $v^2_k$ of single-particle orbitals coupled by octupole interaction in $^{220,222,224,226}$Rn as a function of octupole deformation $\beta_{30}$ calculated with parameter set DD-ME2+S-D1. The quadrupole deformations $\beta_{20}$ of these nuclei are fixed at the ground-state values as given in Table~\ref{tab1}. }
    \label{fig:occupy}
\end{figure*}
In the upper panels of Fig.~\ref{fig:level}, we present the single-neutron levels near the Fermi surface, where neutron orbital pairs with octupole correlations are marked with Nilsson quantum numbers at $\beta_{30}=0$. The splitting between neutron orbitals coupled with octupole interaction becomes larger as the octupole deformation increases. To elucidate the relationship between this level splitting behavior and the emergence of octupole deformation, we employ a schematic model for a qualitative interpretation, as shown in Fig.~\ref{fig:oct}.

We denote the single-particle energies of a pair of octupole-correlated orbitals as $\varepsilon_1$ and $\varepsilon_2$ at $\beta_{30}=0$. %with corresponding occupation probabilities $v^2_1$ and $v^2_2$
The corresponding single-particle Hamiltonian reads
\begin{equation}
    h_{0}=
    \begin{pmatrix}
     \varepsilon_2 & 0 \\
     0   & \varepsilon_1
    \end{pmatrix}.
    \label{1}
\end{equation}
When octupole deformation emerges, the reflection-asymmetric mean field introduces parity mixing in the single-particle eigen-states. This effect can be reflected in the off-diagonal elements $V_{\rm{oct}}$ of the Hamiltonian
\begin{equation}
    h_{\rm{oct}}=
    \begin{pmatrix}
     \varepsilon_2 & V_{\rm{oct}} \\
     V_{\rm{oct}}   & \varepsilon_1
    \end{pmatrix}.
\end{equation}
The corresponding energies become
\begin{equation}
\begin{cases}
\varepsilon^\prime_1 = \frac{\varepsilon_1+\varepsilon_2}{2}+\sqrt{\delta^2+V_{\rm{oct}}^2}, \\
\varepsilon^\prime_2 = \frac{\varepsilon_1+\varepsilon_2}{2}-\sqrt{\delta^2+V_{\rm{oct}}^2},
\end{cases}
\label{2}
\end{equation}
where $\delta=\frac{\varepsilon_1-\varepsilon_2}{2}$. This model clearly explains the splitting behavior of octupole-correlated orbitals under the reflection-asymmetric mean field. 

It can be seen that the reflection-asymmetric mean field does not lead to an overall energy reduction of these two single-particle levels. To explain the energy gain of octupole deformation, the occupation probabilities must be taken into account. The single-particle levels splitting caused by octupole interaction would increase the occupation probability of the lower-energy orbital and decrease that of the higher-energy orbital, resulting in a lower total energy for the reflection-asymmetric configuration. This means that static octupole deformation occurs only when the occupation probabilities of octupole-correlated orbitals near the Fermi surface change significantly with octupole deformation. %Alternatively, while the disturbance from other single-particle states near the Fermi surface suppresses the development of a large shell gap, the resulting level splitting nevertheless greatly reduces the shell correction energy, consequently enhancing the stability of octupole-deformed shapes. 

In Fig.~\ref{fig:occupy}, we present occupation probabilities $v^2_k$ of single-particle orbitals coupled by octupole interaction in $^{220,222,224,226}$Rn as a function of octupole deformation $\beta_{30}$ calculated with parameter set PC-PK+S-D1. It can be seen that only $^{222}$Rn and $^{224}$Rn exhibit substantial changes in the occupation probabilities of relevant orbitals labeled with  $\nu$[642]3/2 and $\nu$[761]3/2 at $\beta_{30}=0$. Another pair of orbitals around Fermi surface labeled with $\nu$[633]5/2 and $\nu$[752]5/2 at $\beta_{30}=0$, do not contribute a lot in octupole energy gain. %This agrees well with our calculation results showing static octupole deformations in $^{222}$Rn and $^{224}$Rn.

\section{summary and perspectives}\label{SP}

In this work, the ground-state properties and PESs of $^{220,222,224,226}$Rn isotopes have been investigated by using the MDC-RMF model. The functionals PC-PK1, DD-PC1, and DD-ME2 and separable pairing force of finite-range are used in the calculations. Predictions are made that octupole minima exist in the PES of $^{222,224}$Rn, but not in $^{220,226}$Rn. The functional- and parameter-independence of the conclusion has been examined. The energy gain due to octupole deformation, ranging from 0.08 to 0.79 MeV, decreases as the pairing strength increases. As the octupole deformation increases, the neutron paring energy decreases and the proton pairing energy increases while there is no obvious change in total pairing energy. Meanwhile, we employed a schematic two-level model to qualitatively discuss the origin of octupole deformation. It is shown that a pair of octupole-driving orbitals splits in the octupole deformed mean field which appears as off-diagonal elements in the single-particle Hamiltonian. Such splitting leads to a change of occupation probabilities of these two orbitals, in particular, an increase of occupation probability of the lower orbital, inducing octupole deformation if the change is large enough. We demonstrate that the octupole deformation in $^{222,224}$Rn originates from the octupole correlations between the $\nu$[642]3/2 (from $\nu g_{9/2}$) and $\nu$[761]3/2 (from $\nu j_{15/2}$) orbitals. Experimental confirmation of the present results is highly desired, such as measurement of the $E3$ transition strength from the ground
state to the first octupole $3^{-}$ state, the occurrence of relatively large $E1$ transition probabilities between the yrast positive- and negative-parity bands. Additionally, the measurement of $B(E3)$ can confirm the systematics of octupole correlations along the $^{220,222,224,226}$Rn isotopic chain, i.e., octupole correlations are stronger in $^{222,224}$Rn than in $^{220,226}$Rn. Further theoretical studies on these nuclei are also called for, particularly in exploring beyond mean field correlations through approaches such as the generator coordinate method.

\begin{acknowledgments}
 Fruitful discussions with Shivani Jain, Bing-Nan Lu, Zhen-Hua Zhang, and Yu-Ting Rong are gratefully acknowledged. 
This work is supported by the National Natural Science Foundation of China (Grant Nos. 12447101, 12347139, 12375118, 12435008, and W2412043), the National Key R\&D Program of China (Grant Nos. 2023YFA1606500 and 2024YFE0109800) and the CAS Strategic Priority Research Program (Grant Nos. XDB34010100 and XDB0920000). The results described in this paper are obtained on the High-performance Computing Cluster of ITP-CAS and the ScGrid of the Supercomputing Center, Computer Network Information Center of Chinese Academy of Sciences. 
\end{acknowledgments}


\bibliography{apssamp.bib}

@article{RevModPhys.68.349,
  title = {Intrinsic reflection asymmetry in atomic nuclei},
  author = {Butler, P. A. and Nazarewicz, W.},
  journal = {Rev. Mod. Phys.},
  volume = {68},
  issue = {2},
  pages = {349--421},
  numpages = {0},
  year = {1996},
  month = {Apr},
  publisher = {American Physical Society},
  doi = {10.1103/RevModPhys.68.349},
  url = {https://link.aps.org/doi/10.1103/RevModPhys.68.349}
}

@article{LANE196039,
title = {On the existence of octupole vibrations in nuclei},
journal = {Nucl. Phys.},
volume = {15},
pages = {39-55},
year = {1960},
issn = {0029-5582},
doi = {https://doi.org/10.1016/0029-5582(60)90280-7},
url = {https://www.sciencedirect.com/science/article/pii/0029558260902807},
author = {A.M. Lane and E.D. Pendlebury}
}

@article{PhysRevLett.124.042503,
  title = {Evolution of Octupole Deformation in Radium Nuclei from Coulomb Excitation of Radioactive $^{222}\mathrm{Ra}$ and $^{228}\mathrm{Ra}$ Beams},
  author = {Butler, P. A. and Gaffney, L. P. and Spagnoletti, P. and Abrahams, K. and Bowry, M. and Cederk\"all, J. and de Angelis, G. and De Witte, H. and Garrett, P. E. and Goldkuhle, A. and Henrich, C. and Illana, A. and Johnston, K. and Joss, D. T. and Keatings, J. M. and Kelly, N. A. and Komorowska, M. and Konki, J. and Kr\"oll, T. and Lozano, M. and Nara Singh, B. S. and O'Donnell, D. and Ojala, J. and Page, R. D. and Pedersen, L. G. and Raison, C. and Reiter, P. and Rodriguez, J. A. and Rosiak, D. and Rothe, S. and Scheck, M. and Seidlitz, M. and Shneidman, T. M. and Siebeck, B. and Sinclair, J. and Smith, J. F. and Stryjczyk, M. and Van Duppen, P. and Vinals, S. and Virtanen, V. and Warr, N. and Wrzosek-Lipska, K. and Zieli\ifmmode \acute{n}\else \'{n}\fi{}ska, M.},
  journal = {Phys. Rev. Lett.},
  volume = {124},
  issue = {4},
  pages = {042503},
  numpages = {6},
  year = {2020},
  month = {Jan},
  publisher = {American Physical Society},
  doi = {10.1103/PhysRevLett.124.042503},
  url = {https://link.aps.org/doi/10.1103/PhysRevLett.124.042503}
}

@article{PhysRevLett.114.233002,
  title = {First Measurement of the Atomic Electric Dipole Moment of $^{225}\mathrm{Ra}$},
  author = {Parker, R. H. and Dietrich, M. R. and Kalita, M. R. and Lemke, N. D. and Bailey, K. G. and Bishof, M. and Greene, J. P. and Holt, R. J. and Korsch, W. and Lu, Z.-T. and Mueller, P. and O'Connor, T. P. and Singh, J. T.},
  journal = {Phys. Rev. Lett.},
  volume = {114},
  issue = {23},
  pages = {233002},
  numpages = {5},
  year = {2015},
  month = {Jun},
  publisher = {American Physical Society},
  doi = {10.1103/PhysRevLett.114.233002},
  url = {https://link.aps.org/doi/10.1103/PhysRevLett.114.233002}
}

@article{PhysRevC.89.014323,
  title = {Multidimensionally-constrained relativistic mean-field models and potential-energy surfaces of actinide nuclei},
  author = {Lu, Bing-Nan and Zhao, Jie and Zhao, En-Guang and Zhou, Shan-Gui},
  journal = {Phys. Rev. C},
  volume = {89},
  issue = {1},
  pages = {014323},
  numpages = {15},
  year = {2014},
  month = {Jan},
  publisher = {American Physical Society},
  doi = {10.1103/PhysRevC.89.014323},
  url = {https://link.aps.org/doi/10.1103/PhysRevC.89.014323}
}

@article{PhysRevC.85.011301,
  title = {Potential energy surfaces of actinide nuclei from a multidimensional constrained covariant density functional theory: Barrier heights and saddle point shapes},
  author = {Lu, Bing-Nan and Zhao, En-Guang and Zhou, Shan-Gui},
  journal = {Phys. Rev. C},
  volume = {85},
  issue = {1},
  pages = {011301},
  numpages = {5},
  year = {2012},
  month = {Jan},
  publisher = {American Physical Society},
  doi = {10.1103/PhysRevC.85.011301},
  url = {https://link.aps.org/doi/10.1103/PhysRevC.85.011301}
}

@article{Nature.497.199, 
author = {Gaffney, Liam and Butler, P and Scheck, Marcus and Hayes, Adam and Wenander, F and Albers, M and Bastin, Beyhan and Bauer, Christopher and Blazhev, A and Bönig, Sabine and Bree, N and Cederkäll, J and Chupp, T. and Cline, Douglas and Cocolios, Thomas and Davinson, T and De Witte, Hilde and Diriken, Jan and Grahn, T and Zielinska, M},
year = {2013},
month = {05},
pages = {199-204},
title = {Studies of pear-shaped nuclei using accelerated radioactive beams},
volume = {497},
journal = {Nature},
doi = {10.1038/nature12073},
url = {https://doi.org/10.1038/nature12073}
}

@article{Nat.Phys.16.853,
author = {Chishti, M. M. R. and O'Donnell, D. and Battaglia, G. and Bowry, M. and Jaroszynski, D. A. and Singh, B. S. Nara and Scheck, M. and Spagnoletti, P. and Smith, J. F.},
year = {2020},
month = {08},
pages = {853},
title = {Direct measurement of the intrinsic electric dipole moment in pear-shaped thorium-228},
volume = {16},
journal = {Nat. Phys.},
doi = {10.1038/s41567-020-0899-4},
url = {https://doi.org/10.1038/s41567-020-0899-4}
}

@article{PhysRevLett.76.4316,
  title = {Collective T- and P-Odd Electromagnetic Moments in Nuclei with Octupole Deformations},
  author = {Auerbach, N. and Flambaum, V. V. and Spevak, V.},
  journal = {Phys. Rev. Lett.},
  volume = {76},
  issue = {23},
  pages = {4316--4319},
  numpages = {0},
  year = {1996},
  month = {Jun},
  publisher = {American Physical Society},
  doi = {10.1103/PhysRevLett.76.4316},
  url = {https://link.aps.org/doi/10.1103/PhysRevLett.76.4316}
}

@article{PhysRevC.56.1357,
  title = {Enhanced $T$-odd, $P$-odd electromagnetic moments in reflection asymmetric nuclei},
  author = {Spevak, V. and Auerbach, N. and Flambaum, V. V.},
  journal = {Phys. Rev. C},
  volume = {56},
  issue = {3},
  pages = {1357--1369},
  numpages = {0},
  year = {1997},
  month = {Sep},
  publisher = {American Physical Society},
  doi = {10.1103/PhysRevC.56.1357},
  url = {https://link.aps.org/doi/10.1103/PhysRevC.56.1357}
}

@article{NAZAREWICZ1984269,
title = {Analysis of octupole instability in medium-mass and heavy nuclei},
journal = {Nucl. Phys. A},
volume = {429},
number = {2},
pages = {269-295},
year = {1984},
issn = {0375-9474},
doi = {https://doi.org/10.1016/0375-9474(84)90208-2},
url = {https://www.sciencedirect.com/science/article/pii/0375947484902082},
author = {W. Nazarewicz and P. Olanders and I. Ragnarsson and J. Dudek and G.A. Leander and P. Möller and E. Ruchowsa}
}

@article{PhysRevC.108.024303,
  title = {Origin of octupole deformation softness in atomic nuclei},
  author = {Minh Loc, Bui and Le Anh, Nguyen and Papakonstantinou, Panagiota and Auerbach, Naftali},
  journal = {Phys. Rev. C},
  volume = {108},
  issue = {2},
  pages = {024303},
  numpages = {6},
  year = {2023},
  month = {Aug},
  publisher = {American Physical Society},
  doi = {10.1103/PhysRevC.108.024303},
  url = {https://link.aps.org/doi/10.1103/PhysRevC.108.024303}
}

@article{PhysRevC.41.R1351,
  title = {Information on octupole deformation from the systematics of B(E3;${0}_{1}^{+}$\ensuremath{\rightarrow}${3}_{1}^{\mathrm{\ensuremath{-}}}$) values},
  author = {Spear, R. H. and Catford, W. N.},
  journal = {Phys. Rev. C},
  volume = {41},
  issue = {4},
  pages = {R1351--R1354},
  numpages = {0},
  year = {1990},
  month = {Apr},
  publisher = {American Physical Society},
  doi = {10.1103/PhysRevC.41.R1351},
  url = {https://link.aps.org/doi/10.1103/PhysRevC.41.R1351}
}

@article{PhysRevC.42.1264,
  title = {New ``Octupole-driving particle numbers'' from examination of ${3}_{1}^{\mathrm{\ensuremath{-}}}$ state energies},
  author = {Cottle, P. D.},
  journal = {Phys. Rev. C},
  volume = {42},
  issue = {4},
  pages = {1264--1266},
  numpages = {0},
  year = {1990},
  month = {Oct},
  publisher = {American Physical Society},
  doi = {10.1103/PhysRevC.42.1264},
  url = {https://link.aps.org/doi/10.1103/PhysRevC.42.1264}
}

@article{PhysRevC.79.064304,
  title = {Heavy-element fission barriers},
  author = {M\"oller, Peter and Sierk, Arnold J. and Ichikawa, Takatoshi and Iwamoto, Akira and Bengtsson, Ragnar and Uhrenholt, Henrik and \AA{}berg, Sven},
  journal = {Phys. Rev. C},
  volume = {79},
  issue = {6},
  pages = {064304},
  numpages = {38},
  year = {2009},
  month = {Jun},
  publisher = {American Physical Society},
  doi = {10.1103/PhysRevC.79.064304},
  url = {https://link.aps.org/doi/10.1103/PhysRevC.79.064304}
}

@article{PhysRevC.95.014303,
  title = {Adiabatic fission barriers in superheavy nuclei},
  author = {Jachimowicz, P. and Kowal, M. and Skalski, J.},
  journal = {Phys. Rev. C},
  volume = {95},
  issue = {1},
  pages = {014303},
  numpages = {20},
  year = {2017},
  month = {Jan},
  publisher = {American Physical Society},
  doi = {10.1103/PhysRevC.95.014303},
  url = {https://link.aps.org/doi/10.1103/PhysRevC.95.014303}
}

@article{Nat.Commun.10.2403,
author = {Butler, P. and Gaffney, Liam and Spagnoletti, Pietro and Konki, J. and Scheck, Marcus and Smith, J. and Abrahams, K. and Bowry, Michael and Cederkäll, J. and Chupp, T. and De Witte, Hilde and Garrett, Paul and Goldkuhle, Alina and Henrich, Corinna and Illana Sison, Andres and Johnston, Karl and Joss, David and Keatings, James and Zielinska, M.},
year = {2019},
month = {06},
pages = {2473},
title = {The observation of vibrating pear-shapes in radon nuclei},
volume = {10},
journal = {Nat. Commun.},
doi = {10.1038/s41467-019-10494-5},
url={https://doi.org/10.1038/s41467-019-10494-5}
}

@article{Butler_2016,
doi = {10.1088/0954-3899/43/7/073002},
url = {https://doi.org/10.1088/0954-3899/43/7/073002},
year = {2016},
month = {jun},
publisher = {IOP Publishing},
volume = {43},
number = {7},
pages = {073002},
author = {Butler, P A},
title = {Octupole collectivity in nuclei},
journal = {J. Phys. G: Nucl. Part. Phys.}
}

@article{MOLLER20161,
title = {Nuclear ground-state masses and deformations: FRDM(2012)},
journal = {At. Data Nucl. Data Tables},
volume = {109-110},
pages = {1-204},
year = {2016},
issn = {0092-640X},
doi = {https://doi.org/10.1016/j.adt.2015.10.002},
url = {https://www.sciencedirect.com/science/article/pii/S0092640X1600005X},
author = {P. Möller and A.J. Sierk and T. Ichikawa and H. Sagawa}
}

@article{PhysRevC.86.054306,
  title = {Electromagnetic transition strengths in soft deformed nuclei},
  author = {Robledo, L. M. and Bertsch, G. F.},
  journal = {Phys. Rev. C},
  volume = {86},
  issue = {5},
  pages = {054306},
  numpages = {7},
  year = {2012},
  month = {Nov},
  publisher = {American Physical Society},
  doi = {10.1103/PhysRevC.86.054306},
  url = {https://link.aps.org/doi/10.1103/PhysRevC.86.054306}
}

@article{PhysRevC.88.051302,
  title = {Quadrupole-octupole coupling in the light actinides},
  author = {Robledo, L. M. and Butler, P. A.},
  journal = {Phys. Rev. C},
  volume = {88},
  issue = {5},
  pages = {051302},
  numpages = {5},
  year = {2013},
  month = {Nov},
  publisher = {American Physical Society},
  doi = {10.1103/PhysRevC.88.051302},
  url = {https://link.aps.org/doi/10.1103/PhysRevC.88.051302}
}

@article{PhysRevC.93.044304,
  title = {Octupole deformation in the ground states of even-even nuclei: A global analysis within the covariant density functional theory},
  author = {Agbemava, S. E. and Afanasjev, A. V. and Ring, P.},
  journal = {Phys. Rev. C},
  volume = {93},
  issue = {4},
  pages = {044304},
  numpages = {24},
  year = {2016},
  month = {Apr},
  publisher = {American Physical Society},
  doi = {10.1103/PhysRevC.93.044304},
  url = {https://link.aps.org/doi/10.1103/PhysRevC.93.044304}
}

@article{Zhou_2016,
doi = {10.1088/0031-8949/91/6/063008},
url = {https://dx.doi.org/10.1088/0031-8949/91/6/063008},
year = {2016},
month = {may},
publisher = {IOP Publishing},
volume = {91},
number = {6},
pages = {063008},
author = {Zhou, Shan-Gui},
title = {Multidimensionally constrained covariant density functional theories—nuclear shapes and potential energy surfaces},
journal = {Phys. Scr.}
}

@article{PhysRevC.91.014321,
  title = {Multidimensionally constrained relativistic mean-field study of triple-humped barriers in actinides},
  author = {Zhao, Jie and Lu, Bing-Nan and Vretenar, Dario and Zhao, En-Guang and Zhou, Shan-Gui},
  journal = {Phys. Rev. C},
  volume = {91},
  issue = {1},
  pages = {014321},
  numpages = {9},
  year = {2015},
  month = {Jan},
  publisher = {American Physical Society},
  doi = {10.1103/PhysRevC.91.014321},
  url = {https://link.aps.org/doi/10.1103/PhysRevC.91.014321}
}

@article{Meng:2019mff,
  title = {Ground State Properties and Potential Energy Surfaces of {{270Hs}} from Multidimensionally-Constrained Relativistic Mean Field Model},
  author = {Meng, Xu and Lu, BingNan and Zhou, ShanGui},
  year = {2020},
  journal = {Sci. China Phys. Mech. Astron.},
  volume = {63},
  number = {1},
  eprinttype = {arXiv},
  eprintclass = {nucl-th},
  pages = {212011},
  issn = {1869-1927},
  doi = {10.1007/s11433-019-9422-1},
  url = {https://doi.org/10.1007/s11433-019-9422-1}
}

@article{PhysRevC.95.014320,
  title = {Tetrahedral shapes of neutron-rich Zr isotopes from a multidimensionally constrained relativistic Hartree-Bogoliubov model},
  author = {Zhao, Jie and Lu, Bing-Nan and Zhao, En-Guang and Zhou, Shan-Gui},
  journal = {Phys. Rev. C},
  volume = {95},
  issue = {1},
  pages = {014320},
  numpages = {16},
  year = {2017},
  month = {Jan},
  publisher = {American Physical Society},
  doi = {10.1103/PhysRevC.95.014320},
  url = {https://link.aps.org/doi/10.1103/PhysRevC.95.014320}
}

@article{PhysRevC.86.057304,
  title = {Nonaxial-octupole ${Y}_{32}$ correlations in $N=150$ isotones from multidimensional constrained covariant density functional theories},
  author = {Zhao, Jie and Lu, Bing-Nan and Zhao, En-Guang and Zhou, Shan-Gui},
  journal = {Phys. Rev. C},
  volume = {86},
  issue = {5},
  pages = {057304},
  numpages = {4},
  year = {2012},
  month = {Nov},
  publisher = {American Physical Society},
  doi = {10.1103/PhysRevC.86.057304},
  url = {https://link.aps.org/doi/10.1103/PhysRevC.86.057304}
}

@article{PhysRevC.84.014328,
  title = {Quadrupole deformation $(\ensuremath{\beta},\ensuremath{\gamma})$ of light $\ensuremath{\Lambda}$ hypernuclei in a constrained relativistic mean field model: Shape evolution and shape polarization effect of the $\ensuremath{\Lambda}$ hyperon},
  author = {Lu, Bing-Nan and Zhao, En-Guang and Zhou, Shan-Gui},
  journal = {Phys. Rev. C},
  volume = {84},
  issue = {1},
  pages = {014328},
  numpages = {10},
  year = {2011},
  month = {Jul},
  publisher = {American Physical Society},
  doi = {10.1103/PhysRevC.84.014328},
  url = {https://link.aps.org/doi/10.1103/PhysRevC.84.014328}
}

@article{PhysRevC.89.044307,
  title = {Superdeformed $\ensuremath{\Lambda}$ hypernuclei within relativistic mean field models},
  author = {Lu, Bing-Nan and Hiyama, Emiko and Sagawa, Hiroyuki and Zhou, Shan-Gui},
  journal = {Phys. Rev. C},
  volume = {89},
  issue = {4},
  pages = {044307},
  numpages = {7},
  year = {2014},
  month = {Apr},
  publisher = {American Physical Society},
  doi = {10.1103/PhysRevC.89.044307},
  url = {https://link.aps.org/doi/10.1103/PhysRevC.89.044307}
}

@article{RONG2020135533,
title = {Strength of pairing interaction for hyperons in multistrangeness hypernuclei},
journal = {Phys. Lett. B},
volume = {807},
pages = {135533},
year = {2020},
issn = {0370-2693},
doi = {https://doi.org/10.1016/j.physletb.2020.135533},
url = {https://www.sciencedirect.com/science/article/pii/S0370269320303373},
author = {Yu-Ting Rong and Pengwei Zhao and Shan-Gui Zhou}
}

@article{Chen:2021kde,
  title = {Possible Shape Coexistence in {{Ne}} Isotopes and the Impurity Effect of {{Λ}} Hyperon},
  author = {Chen, Chen and Sun, Qian-Kun and Li, Yu-Xiao and Sun, Ting-Ting},
  date = {2021-07-01},
  year = {2021},
  journal = {Sci. China Phys. Mech. Astron.},
  shortjournal = {Sci. China Phys. Mech. Astron.},
  volume = {64},
  number = {8},
  pages = {282011},
  issn = {1869-1927},
  doi = {10.1007/s11433-021-1721-1},
  url = {https://doi.org/10.1007/s11433-021-1721-1}
}

@article{Sun_2022,
doi = {10.1088/1674-1137/ac6153},
url = {https://dx.doi.org/10.1088/1674-1137/ac6153},
year = {2022},
month = {jul},
publisher = {Chinese Physical Society and the Institute of High Energy Physics of the Chinese Academy of Sciences and the Institute of Modern Physics of the Chinese Academy of Sciences and IOP Publishing Ltd},
volume = {46},
number = {7},
pages = {074106},
author = {Sun, Qian-Kun and Sun, Ting-Ting and Zhang, Wei and Zhang, Shi-Sheng and Chen, Chen},
title = {Possible shape coexistence in odd-A Ne isotopes and the impurity effects of Λ hyperons *},
journal = {Chin. Phys. C}
}

@article{Reinhard_1989,
doi = {10.1088/0034-4885/52/4/002},
url = {https://dx.doi.org/10.1088/0034-4885/52/4/002},
year = {1989},
month = {apr},
publisher = {},
volume = {52},
number = {4},
pages = {439},
author = {P -G Reinhard},
title = {The relativistic mean-field description of nuclei and nuclear dynamics},
journal = {Rep. Prog. Phys.}
}

@article{RING1996193,
title = {Relativistic mean field theory in finite nuclei},
journal = {Prog. Part. Nucl. Phys.},
volume = {37},
pages = {193-263},
year = {1996},
issn = {0146-6410},
doi = {https://doi.org/10.1016/0146-6410(96)00054-3},
url = {https://www.sciencedirect.com/science/article/pii/0146641096000543},
author = {P. Ring}
}

@article{VRETENAR2005101,
title = {Relativistic Hartree–Bogoliubov theory: static and dynamic aspects of exotic nuclear structure},
journal = {Phys. Rep.},
volume = {409},
number = {3},
pages = {101-259},
year = {2005},
issn = {0370-1573},
doi = {https://doi.org/10.1016/j.physrep.2004.10.001},
url = {https://www.sciencedirect.com/science/article/pii/S0370157304004545},
author = {D. Vretenar and A.V. Afanasjev and G.A. Lalazissis and P. Ring}
}

@article{MENG2006470,
title = {Relativistic continuum Hartree Bogoliubov theory for ground-state properties of exotic nuclei},
journal = {Prog. Part. Nucl. Phys.},
volume = {57},
number = {2},
pages = {470-563},
year = {2006},
issn = {0146-6410},
doi = {https://doi.org/10.1016/j.ppnp.2005.06.001},
url = {https://www.sciencedirect.com/science/article/pii/S014664100500075X},
author = {J. Meng and H. Toki and S.G. Zhou and S.Q. Zhang and W.H. Long and L.S. Geng}
}

@article{Paar_2007,
doi = {10.1088/0034-4885/70/5/R02},
url = {https://dx.doi.org/10.1088/0034-4885/70/5/R02},
year = {2007},
month = {may},
publisher = {},
volume = {70},
number = {5},
pages = {R02},
author = {Paar, Nils and Vretenar, Dario and Khan, Elias and Colò, Gianluca},
title = {Exotic modes of excitation in atomic nuclei far from stability},
journal = {Rep. Prog. Phys.}
}

@article{NIKSIC2011519,
title = {Relativistic nuclear energy density functionals: Mean-field and beyond},
journal = {Prog. Part. Nucl. Phys.},
volume = {66},
number = {3},
pages = {519-548},
year = {2011},
issn = {0146-6410},
doi = {https://doi.org/10.1016/j.ppnp.2011.01.055},
url = {https://www.sciencedirect.com/science/article/pii/S0146641011000561},
author = {T. Nikšić and D. Vretenar and P. Ring}
}

@article{PhysRevC.46.1757,
  title = {Nuclear ground state properties in a relativistic point coupling model},
  author = {Nikolaus, B. A. and Hoch, T. and Madland, D. G.},
  journal = {Phys. Rev. C},
  volume = {46},
  issue = {5},
  pages = {1757--1781},
  numpages = {0},
  year = {1992},
  month = {Nov},
  publisher = {American Physical Society},
  doi = {10.1103/PhysRevC.46.1757},
  url = {https://link.aps.org/doi/10.1103/PhysRevC.46.1757}
}

@article{PhysRevC.65.044308,
  title = {Nuclear ground state observables and QCD scaling in a refined relativistic point coupling model},
  author = {B\"urvenich, T. and Madland, D. G. and Maruhn, J. A. and Reinhard, P.-G.},
  journal = {Phys. Rev. C},
  volume = {65},
  issue = {4},
  pages = {044308},
  numpages = {23},
  year = {2002},
  month = {Mar},
  publisher = {American Physical Society},
  doi = {10.1103/PhysRevC.65.044308},
  url = {https://link.aps.org/doi/10.1103/PhysRevC.65.044308}
}

@article{PhysRevLett.68.3408,
  title = {Relativistic density-dependent Hartree approach for finite nuclei},
  author = {Brockmann, R. and Toki, H.},
  journal = {Phys. Rev. Lett.},
  volume = {68},
  issue = {23},
  pages = {3408--3411},
  numpages = {0},
  year = {1992},
  month = {Jun},
  publisher = {American Physical Society},
  doi = {10.1103/PhysRevLett.68.3408},
  url = {https://link.aps.org/doi/10.1103/PhysRevLett.68.3408}
}

@article{PhysRevC.80.024313,
  title = {Axially deformed relativistic Hartree Bogoliubov theory with a separable pairing force},
  author = {Tian, Yuan and Ma, Zhong-yu and Ring, P.},
  journal = {Phys. Rev. C},
  volume = {80},
  issue = {2},
  pages = {024313},
  numpages = {7},
  year = {2009},
  month = {Aug},
  publisher = {American Physical Society},
  doi = {10.1103/PhysRevC.80.024313},
  url = {https://link.aps.org/doi/10.1103/PhysRevC.80.024313}
}

@article{TIAN200944,
title = {A finite range pairing force for density functional theory in superfluid nuclei},
journal = {Phys. Lett. B},
volume = {676},
number = {1},
pages = {44-50},
year = {2009},
issn = {0370-2693},
doi = {https://doi.org/10.1016/j.physletb.2009.04.067},
url = {https://www.sciencedirect.com/science/article/pii/S0370269309004912},
author = {Y. Tian and Z.Y. Ma and P. Ring}
}

@article{RING199777,
title = {Computer program for the relativistic mean field description of the ground state properties of even-even axially deformed nuclei},
journal = {Comput. Phys. Commun.},
volume = {105},
number = {1},
pages = {77-97},
year = {1997},
issn = {0010-4655},
doi = {https://doi.org/10.1016/S0010-4655(97)00022-2},
url = {https://www.sciencedirect.com/science/article/pii/S0010465597000222},
author = {P. Ring and Y.K. Gambhir and G.A. Lalazissis}
}

@article{PhysRevC.78.034318,
  title = {Relativistic nuclear energy density functionals: Adjusting parameters to binding energies},
  author = {Nik\ifmmode \check{s}\else \v{s}\fi{}i\ifmmode \acute{c}\else \'{c}\fi{}, T. and Vretenar, D. and Ring, P.},
  journal = {Phys. Rev. C},
  volume = {78},
  issue = {3},
  pages = {034318},
  numpages = {19},
  year = {2008},
  month = {Sep},
  publisher = {American Physical Society},
  doi = {10.1103/PhysRevC.78.034318},
  url = {https://link.aps.org/doi/10.1103/PhysRevC.78.034318}
}

@article{PhysRevC.82.054319,
  title = {New parametrization for the nuclear covariant energy density functional with a point-coupling interaction},
  author = {Zhao, P. W. and Li, Z. P. and Yao, J. M. and Meng, J.},
  journal = {Phys. Rev. C},
  volume = {82},
  issue = {5},
  pages = {054319},
  numpages = {14},
  year = {2010},
  month = {Nov},
  publisher = {American Physical Society},
  doi = {10.1103/PhysRevC.82.054319},
  url = {https://link.aps.org/doi/10.1103/PhysRevC.82.054319}
}

@article{PhysRevC.71.024312,
  title = {New relativistic mean-field interaction with density-dependent meson-nucleon couplings},
  author = {Lalazissis, G. A. and Nik\ifmmode \check{s}\else \v{s}\fi{}i\ifmmode \acute{c}\else \'{c}\fi{}, T. and Vretenar, D. and Ring, P.},
  journal = {Phys. Rev. C},
  volume = {71},
  issue = {2},
  pages = {024312},
  numpages = {10},
  year = {2005},
  month = {Feb},
  publisher = {American Physical Society},
  doi = {10.1103/PhysRevC.71.024312},
  url = {https://link.aps.org/doi/10.1103/PhysRevC.71.024312}
}

@article{Huang_2021,
doi = {10.1088/1674-1137/abddb0},
url = {https://dx.doi.org/10.1088/1674-1137/abddb0},
year = {2021},
month = {mar},
publisher = {Chinese Physical Society and the Institute of High Energy Physics of the Chinese Academy of Sciences and the Institute of Modern Physics of the Chinese Academy of Sciences and IOP Publishing Ltd},
volume = {45},
number = {3},
pages = {030002},
author = {Huang, W.J. and Wang, Meng and Kondev, F.G. and Audi, G. and Naimi, S.},
title = {The AME 2020 atomic mass evaluation (I). Evaluation of input data, and adjustment procedures*},
journal = {Chin. Phys. C}
}

@article{Wang_2021,
doi = {10.1088/1674-1137/abddaf},
url = {https://dx.doi.org/10.1088/1674-1137/abddaf},
year = {2021},
month = {mar},
publisher = {Chinese Physical Society and the Institute of High Energy Physics of the Chinese Academy of Sciences and the Institute of Modern Physics of the Chinese Academy of Sciences and IOP Publishing Ltd},
volume = {45},
number = {3},
pages = {030003},
author = {Wang, Meng and Huang, W.J. and Kondev, F.G. and Audi, G. and Naimi, S.},
title = {The AME 2020 atomic mass evaluation (II). Tables, graphs and references*},
journal = {Chin. Phys. C}
}

@article{PhysRevLett.116.112503,
  title = {Direct Evidence of Octupole Deformation in Neutron-Rich $^{144}\mathrm{Ba}$},
  author = {Bucher, B. and Zhu, S. and Wu, C. Y. and Janssens, R. V. F. and Cline, D. and Hayes, A. B. and Albers, M. and Ayangeakaa, A. D. and Butler, P. A. and Campbell, C. M. and Carpenter, M. P. and Chiara, C. J. and Clark, J. A. and Crawford, H. L. and Cromaz, M. and David, H. M. and Dickerson, C. and Gregor, E. T. and Harker, J. and Hoffman, C. R. and Kay, B. P. and Kondev, F. G. and Korichi, A. and Lauritsen, T. and Macchiavelli, A. O. and Pardo, R. C. and Richard, A. and Riley, M. A. and Savard, G. and Scheck, M. and Seweryniak, D. and Smith, M. K. and Vondrasek, R. and Wiens, A.},
  journal = {Phys. Rev. Lett.},
  volume = {116},
  issue = {11},
  pages = {112503},
  numpages = {5},
  year = {2016},
  month = {Mar},
  publisher = {American Physical Society},
  doi = {10.1103/PhysRevLett.116.112503},
  url = {https://link.aps.org/doi/10.1103/PhysRevLett.116.112503}
}

@Book{Bohr1969_Nucl_Structure,
  author = {Bohr, A. and Mottelson, B. R.},
  title = {Nuclear Structure Vol 2},
  year = {1969},
  publisher    = {New York: Benjamin Inc}
}

@article{PhysRevLett.116.112501,
  title = {Evidence for Octupole Correlations in Multiple Chiral Doublet Bands},
  author = {Liu, C. and Wang, S. Y. and Bark, R. A. and Zhang, S. Q. and Meng, J. and Qi, B. and Jones, P. and Wyngaardt, S. M. and Zhao, J. and Xu, C. and Zhou, S.-G. and Wang, S. and Sun, D. P. and Liu, L. and Li, Z. Q. and Zhang, N. B. and Jia, H. and Li, X. Q. and Hua, H. and Chen, Q. B. and Xiao, Z. G. and Li, H. J. and Zhu, L. H. and Bucher, T. D. and Dinoko, T. and Easton, J. and Juh\'asz, K. and Kamblawe, A. and Khaleel, E. and Khumalo, N. and Lawrie, E. A. and Lawrie, J. J. and Majola, S. N. T. and Mullins, S. M. and Murray, S. and Ndayishimye, J. and Negi, D. and Noncolela, S. P. and Ntshangase, S. S. and Nyak\'o, B. M. and Orce, J. N. and Papka, P. and Sharpey-Schafer, J. F. and Shirinda, O. and Sithole, P. and Stankiewicz, M. A. and Wiedeking, M.},
  journal = {Phys. Rev. Lett.},
  volume = {116},
  issue = {11},
  pages = {112501},
  numpages = {6},
  year = {2016},
  month = {Mar},
  publisher = {American Physical Society},
  doi = {10.1103/PhysRevLett.116.112501},
  url = {https://link.aps.org/doi/10.1103/PhysRevLett.116.112501}
}

@article{Wang_2022,
doi = {10.1088/1674-1137/ac3904},
url = {https://dx.doi.org/10.1088/1674-1137/ac3904},
year = {2022},
month = {feb},
publisher = {Chinese Physical Society and the Institute of High Energy Physics of the Chinese Academy of Sciences and the Institute of Modern Physics of the Chinese Academy of Sciences and IOP Publishing Ltd},
volume = {46},
number = {2},
pages = {024107},
author = {Wang, Xiao-Qian and Sun, Xiang-Xiang and Zhou, Shan-Gui},
title = {Microscopic study of higher-order deformation effects on the ground states of superheavy nuclei around 270Hs *},
journal = {Chin. Phys. C}
}

@article{PhysRevC.95.034329,
  title = {Effect of non-axial octupole shapes in heavy and superheavy nuclei},
  author = {Jachimowicz, P. and Kowal, M. and Skalski, J.},
  journal = {Phys. Rev. C},
  volume = {95},
  issue = {3},
  pages = {034329},
  numpages = {7},
  year = {2017},
  month = {Mar},
  publisher = {American Physical Society},
  doi = {10.1103/PhysRevC.95.034329},
  url = {https://link.aps.org/doi/10.1103/PhysRevC.95.034329}
}

@article{PhysRevC.84.054302,
  title = {Global systematics of octupole excitations in even-even nuclei},
  author = {Robledo, L. M. and Bertsch, G. F.},
  journal = {Phys. Rev. C},
  volume = {84},
  issue = {5},
  pages = {054302},
  numpages = {11},
  year = {2011},
  month = {Nov},
  publisher = {American Physical Society},
  doi = {10.1103/PhysRevC.84.054302},
  url = {https://link.aps.org/doi/10.1103/PhysRevC.84.054302}
}

@article{PhysRevC.102.024311,
  title = {Landscape of pear-shaped even-even nuclei},
  author = {Cao, Yuchen and Agbemava, S. E. and Afanasjev, A. V. and Nazarewicz, W. and Olsen, E.},
  journal = {Phys. Rev. C},
  volume = {102},
  issue = {2},
  pages = {024311},
  numpages = {12},
  year = {2020},
  month = {Aug},
  publisher = {American Physical Society},
  doi = {10.1103/PhysRevC.102.024311},
  url = {https://link.aps.org/doi/10.1103/PhysRevC.102.024311}
}

@article{Buck2008NegativePB,
  title={Negative parity bands in even–even isotopes of Ra, Th, U and Pu},
  author={B. Buck and A. C. Merchant and Sandra M. Perez},
  journal={J. Phys. G: Nucl. Part. Phys.},
  year={2008},
  volume={35},
  pages={085101},
  url={https://api.semanticscholar.org/CorpusID:121087145}
}

@article{PhysRevC.67.014313,
  title = {Cluster interpretation of properties of alternating parity bands in heavy nuclei},
  author = {Shneidman, T. M. and Adamian, G. G. and Antonenko, N. V. and Jolos, R. V. and Scheid, W.},
  journal = {Phys. Rev. C},
  volume = {67},
  issue = {1},
  pages = {014313},
  numpages = {12},
  year = {2003},
  month = {Jan},
  publisher = {American Physical Society},
  doi = {10.1103/PhysRevC.67.014313},
  url = {https://link.aps.org/doi/10.1103/PhysRevC.67.014313}
}

@article{Robledo_2012,
doi = {10.1088/0954-3899/39/10/105103},
url = {https://dx.doi.org/10.1088/0954-3899/39/10/105103},
year = {2012},
month = {aug},
publisher = {IOP Publishing},
volume = {39},
number = {10},
pages = {105103},
author = {Robledo, L M and Rodríguez-Guzmán, R R},
title = {Octupole deformation properties of actinide isotopes within a mean-field approach},
journal = {J. Phys. G: Nucl. Part. Phys.}
}

@article{WANG2019454,
title = {Multiple chiral doublet bands with octupole correlations in reflection-asymmetric triaxial particle rotor model},
journal = {Phys. Lett. B},
volume = {792},
pages = {454-460},
year = {2019},
issn = {0370-2693},
doi = {https://doi.org/10.1016/j.physletb.2019.04.014},
url = {https://www.sciencedirect.com/science/article/pii/S0370269319302461},
author = {Y.Y. Wang and S.Q. Zhang and P.W. Zhao and J. Meng}
}

@article{PhysRevC.102.064328,
  title = {Alternating-parity bands of $^{236,238}\mathrm{U}$ and $^{238,240}\mathrm{Pu}$ in a particle-number-conserving method based on the cranked shell model},
  author = {He, Xiao-Tao and Li, Yu-Chun},
  journal = {Phys. Rev. C},
  volume = {102},
  issue = {6},
  pages = {064328},
  numpages = {11},
  year = {2020},
  month = {Dec},
  publisher = {American Physical Society},
  doi = {10.1103/PhysRevC.102.064328},
  url = {https://link.aps.org/doi/10.1103/PhysRevC.102.064328}
}

@article{LU2025139620,
title = {Octupole correlations in superdeformed bands of 56Ni},
journal = {Phys. Lett. B},
volume = {868},
pages = {139620},
year = {2025},
issn = {0370-2693},
doi = {https://doi.org/10.1016/j.physletb.2025.139620},
url = {https://www.sciencedirect.com/science/article/pii/S0370269325003818},
author = {Xiao Lu and Shivani Jain and Hiroyuki Sagawa and Shan-Gui Zhou}
}

@article{LI2013866,
title = {Simultaneous quadrupole and octupole shape phase transitions in Thorium},
journal = {Phys. Lett. B},
volume = {726},
number = {4},
pages = {866-869},
year = {2013},
issn = {0370-2693},
doi = {https://doi.org/10.1016/j.physletb.2013.09.035},
url = {https://www.sciencedirect.com/science/article/pii/S0370269313007648},
author = {Z.P. Li and B.Y. Song and J.M. Yao and D. Vretenar and J. Meng}
}

@article{PhysRevC.91.014317,
  title = {Octupole bands and simplex inversion in the neutron-rich nucleus $^{145}\mathrm{Ba}$},
  author = {Chen, Yong-Jing and Gao, Zao-Chun and Chen, Yong-Shou and Tu, Ya},
  journal = {Phys. Rev. C},
  volume = {91},
  issue = {1},
  pages = {014317},
  numpages = {7},
  year = {2015},
  month = {Jan},
  publisher = {American Physical Society},
  doi = {10.1103/PhysRevC.91.014317},
  url = {https://link.aps.org/doi/10.1103/PhysRevC.91.014317}
}

@article{Robledo_2015,
doi = {10.1088/0954-3899/42/5/055109},
url = {https://dx.doi.org/10.1088/0954-3899/42/5/055109},
year = {2015},
month = {mar},
publisher = {IOP Publishing},
volume = {42},
number = {5},
pages = {055109},
author = {Robledo, L M},
title = {Ground state octupole correlation energy with effective forces},
journal = {J. Phys. G: Nucl. Part. Phys.}
}

@article{RONG2023137896,
title = {Anatomy of octupole correlations in 96Zr with a symmetry-restored multidimensionally-constrained covariant density functional theory},
journal = {Phys. Lett. B},
volume = {840},
pages = {137896},
year = {2023},
issn = {0370-2693},
doi = {https://doi.org/10.1016/j.physletb.2023.137896},
url = {https://www.sciencedirect.com/science/article/pii/S0370269323002307},
author = {Yu-Ting Rong and Xian-Ye Wu and Bing-Nan Lu and Jiang-Ming Yao}
}

@article{XU2024138893,
title = {Coexistence of pure octupole shapes in the superheavy nucleus 286No},
journal = {Phys. Lett. B},
volume = {856},
pages = {138893},
year = {2024},
issn = {0370-2693},
doi = {https://doi.org/10.1016/j.physletb.2024.138893},
url = {https://www.sciencedirect.com/science/article/pii/S0370269324004519},
author = {F.F. Xu and B. Li and P. Ring and P.W. Zhao}
}

@article{ENGEL198761,
title = {Interacting boson model of collective octupole states: (I). The rotational limit},
journal = {Nucl. Phys. A},
volume = {472},
number = {1},
pages = {61-84},
year = {1987},
issn = {0375-9474},
doi = {https://doi.org/10.1016/0375-9474(87)90220-X},
url = {https://www.sciencedirect.com/science/article/pii/037594748790220X},
author = {J. Engel and F. Iachello}
}

@article{PhysRevC.58.1500,
  title = {Octupole states in deformed actinide nuclei with the interacting boson approximation},
  author = {Cottle, P. D. and Zamfir, N. V.},
  journal = {Phys. Rev. C},
  volume = {58},
  issue = {3},
  pages = {1500--1514},
  numpages = {0},
  year = {1998},
  month = {Sep},
  publisher = {American Physical Society},
  doi = {10.1103/PhysRevC.58.1500},
  url = {https://link.aps.org/doi/10.1103/PhysRevC.58.1500}
}

@article{PhysRevC.67.014305,
  title = {Octupole correlations in U and Pu nuclei},
  author = {Zamfir, N. V. and Kusnezov, Dimitri},
  journal = {Phys. Rev. C},
  volume = {67},
  issue = {1},
  pages = {014305},
  numpages = {8},
  year = {2003},
  month = {Jan},
  publisher = {American Physical Society},
  doi = {10.1103/PhysRevC.67.014305},
  url = {https://link.aps.org/doi/10.1103/PhysRevC.67.014305}
}

@article{serot1986,
  title = { The relativistic nuclear many-body problem},
  author = { B. D. Serot and J. D. Walecka},
  journal = { Adv. Nucl. Phys.},
  volume = {16},
  
  pages = {1},
  
  year = {1986},
  month = {Jan},
}

@article{NAZAREWICZ1985420,
title = {Rotational consequences of stable octupole deformation in nuclei},
journal = {Nucl. Phys. A},
volume = {441},
number = {3},
pages = {420-444},
year = {1985},
issn = {0375-9474},
doi = {https://doi.org/10.1016/0375-9474(85)90154-X},
url = {https://www.sciencedirect.com/science/article/pii/037594748590154X},
author = {W. Nazarewicz and P. Olanders}
}

@article{PhysRevC.63.014314,
  title = {Reflection asymmetric shell model for octupole-deformed nuclei},
  author = {Chen, Y. S. and Gao, Z. C.},
  journal = {Phys. Rev. C},
  volume = {63},
  issue = {1},
  pages = {014314},
  numpages = {7},
  year = {2000},
  month = {Dec},
  publisher = {American Physical Society},
  doi = {10.1103/PhysRevC.63.014314},
  url = {https://link.aps.org/doi/10.1103/PhysRevC.63.014314}
}

@article{doi:10.1142/S0218301323400037,
author = {Han, X. C. and Liu, C. and Wang, S. Y.},
title = {Experimental review of octupole correlations in the A≈80 mass region},
journal = {Int. J. Mod. Phys. E},
volume = {32},
number = {10},
pages = {2340003},
year = {2023},
doi = {10.1142/S0218301323400037},
URL = {https://doi.org/10.1142/S0218301323400037},

}

@article{PhysRevC.85.025802,
  title = {Fission properties for $r$-process nuclei},
  author = {Erler, J. and Langanke, K. and Loens, H. P. and Mart\'{\i}nez-Pinedo, G. and Reinhard, P.-G.},
  journal = {Phys. Rev. C},
  volume = {85},
  issue = {2},
  pages = {025802},
  numpages = {11},
  year = {2012},
  month = {Feb},
  publisher = {American Physical Society},
  doi = {10.1103/PhysRevC.85.025802},
  url = {https://link.aps.org/doi/10.1103/PhysRevC.85.025802}
}

@article{Ebata_2017,
doi = {10.1088/1402-4896/aa6c4c},
url = {https://dx.doi.org/10.1088/1402-4896/aa6c4c},
year = {2017},
month = {may},
publisher = {IOP Publishing},
volume = {92},
number = {6},
pages = {064005},
author = {Ebata, Shuichiro and Nakatsukasa, Takashi},
title = {Octupole deformation in the nuclear chart based on the 3D Skyrme Hartree–Fock plus BCS model},
journal = {Phys. Scr.}
}

@article{LEANDER1984375,
title = {Intrinsic reflection asymmetry in odd-A nuclei},
journal = {Nucl. Phys. A},
volume = {413},
number = {3},
pages = {375-415},
year = {1984},
issn = {0375-9474},
doi = {https://doi.org/10.1016/0375-9474(84)90417-2},
url = {https://www.sciencedirect.com/science/article/pii/0375947484904172},
author = {G.A. Leander and R.K. Sheline}
}

@article{Wang_2022_c,
doi = {10.1088/1572-9494/ac3999},
url = {https://dx.doi.org/10.1088/1572-9494/ac3999},
year = {2022},
month = {jan},
publisher = {IOP Publishing},
volume = {74},
number = {1},
pages = {015303},
author = {Wang, Kun and Lu, Bing-Nan},
title = {The angular momentum and parity projected multidimensionally constrained relativistic Hartree–Bogoliubov model},
journal = { Commun.
 Theor. Phys.}
}

@article{TianYuan_2006,
doi = {10.1088/0256-307X/23/12/029},
url = {https://dx.doi.org/10.1088/0256-307X/23/12/029},
year = {2006},
month = {dec},
publisher = {},
volume = {23},
number = {12},
pages = {3226},
author = {Tian, Yuan and Ma, Zhong-yu and Ring, Peter},
title = {A Separable Pairing Force in Nuclear Matter},
journal = {Chin. Phys. Lett.}
}

@article{PhysRevC.79.064301,
  title = {Separable pairing force for relativistic quasiparticle random-phase approximation},
  author = {Tian, Yuan and Ma, Zhong-yu and Ring, Peter},
  journal = {Phys. Rev. C},
  volume = {79},
  issue = {6},
  pages = {064301},
  numpages = {7},
  year = {2009},
  month = {Jun},
  publisher = {American Physical Society},
  doi = {10.1103/PhysRevC.79.064301},
  url = {https://link.aps.org/doi/10.1103/PhysRevC.79.064301}
}

@article{PhysRevC.86.064324,
  title = {Crucial test for covariant density functional theory with new and accurate mass measurements from Sn to Pa},
  author = {Zhao, P. W. and Song, L. S. and Sun, B. and Geissel, H. and Meng, J.},
  journal = {Phys. Rev. C},
  volume = {86},
  issue = {6},
  pages = {064324},
  numpages = {6},
  year = {2012},
  month = {Dec},
  publisher = {American Physical Society},
  doi = {10.1103/PhysRevC.86.064324},
  url = {https://link.aps.org/doi/10.1103/PhysRevC.86.064324}
}

@article{PhysRevC.92.034302,
  title = {Cluster approach to the structure of $^{240}\mathrm{Pu}$},
  author = {Shneidman, T. M. and Adamian, G. G. and Antonenko, N. V. and Jolos, R. V. and Zhou, Shan-Gui},
  journal = {Phys. Rev. C},
  volume = {92},
  issue = {3},
  pages = {034302},
  numpages = {12},
  year = {2015},
  month = {Sep},
  publisher = {American Physical Society},
  doi = {10.1103/PhysRevC.92.034302},
  url = {https://link.aps.org/doi/10.1103/PhysRevC.92.034302}
}

@article{GAMBHIR1990132,
title = {Relativistic mean field theory for finite nuclei},
journal = {Ann. Phys.},
volume = {198},
number = {1},
pages = {132-179},
year = {1990},
issn = {0003-4916},
doi = {https://doi.org/10.1016/0003-4916(90)90330-Q},
url = {https://www.sciencedirect.com/science/article/pii/000349169090330Q},
author = {Y.K Gambhir and P Ring and A Thimet}
}
\end{document}